\documentclass[11pt]{article}
\usepackage{epsfig}
\usepackage{amsthm}
\usepackage{amssymb}
\usepackage{amsmath}
\usepackage{amsfonts}
\usepackage{color}
\newcommand{\field}[1]{\mathbb{#1}}
\newcommand{\R}{\field{R}}

\newcommand{\N}{\field{N}}
\newcommand{\Z}{\field{Z}}

\newcommand{\E}{\field{E}}
\def\qed{\hfill$\diamondsuit$}

\theoremstyle{example} \theoremstyle{remark} \theoremstyle{lemma}
\theoremstyle{definition} \theoremstyle{corol}
\theoremstyle{proposition} \theoremstyle{condition}
\theoremstyle{assumption}
\newtheorem{assumption}{\n{Assumption}}[section]
\newtheorem{theorem}{\n{Theorem}}[section]

\newtheorem{remark}{\n{Remark}}[section]
\newtheorem{lemma}{\n{Lemma}}[section]

\font\n=cmcsc10
\def\cov{{\mbox{cov}}}
\def\var{{\mbox{var}}}
\def\cum{{\mbox{cum}}}

\begin{document}

\begin{center}
 {\bf\large NONSTATIONARITY-EXTENDED WHITTLE ESTIMATION \footnote{I would like to
 thank Paul Harder for his excellent research assistance on the simulation studies. I am also grateful to three referees and Professor Peter Phillips for constructive
 comments that led to improvement of the paper. The work is  supported in part by NSF grant DMS-0804937.
 Address correspondence to: Xiaofeng Shao, Department of
Statistics, University of Illinois at Urbana-Champaign,  725 South
Wright St, Champaign, IL, 61820; e-mail: xshao@uiuc.edu}}
\end{center}

 \centerline{\textsc{By Xiaofeng Shao}}
\centerline{\today} \centerline {\it University of Illinois at
Urbana-Champaign}

\bigskip

\pagenumbering{arabic}

\setcounter{page}{1}

\noindent  For long memory time series models with uncorrelated but
dependent errors, we establish the asymptotic normality of
 the Whittle estimator under mild conditions. Our framework includes the widely used
 FARIMA models with GARCH-type innovations.   To cover nonstationary fractionally integrated processes,
we extend the idea of Abadir, Distaso and Giraitis (2007, {\it Journal of Econometrics} 141, 1353-1384)
 and develop the nonstationarity-extended Whittle estimation. The resulting
estimator is shown to be asymptotically normal and is more
efficient than the tapered Whittle estimator. Finally, the results from a small simulation study are presented to
corroborate our theoretical findings.

\section{Introduction}

In the recent two decades, there has been a great deal of research
on long memory time series [see Doukhan et al. (2003), Robinson
(2003)]. To model the long memory phenomenon, a widely used model
is the FARIMA$(p,d,q)$ (fractional autoregressive integrated
moving average) model described as follows:
\begin{eqnarray}
\label{eq:farima} \phi(B)(1-B)^{d_X} (X_t-\mu)=\psi(B)u_t,
\end{eqnarray}
where $\mu$ is the mean, $d_X\in (-1/2,1/2)$ is the long memory
parameter, $B$ is the backward shift operator and
$\phi(B)=1-\sum_{i=1}^{p}\phi_i B^i$, $\psi(B)=1+\sum_{i=1}^{q}\psi_i
B^i$ are AR (autoregressive) and MA (moving average) polynomials
respectively.
We call the process $\{X_t\}$ to be fractionally integrated with order
$d_X$, denoted as $X_t\sim I(d_X)$. Typically $\{u_t\}_{t\in\Z}$ are
assumed to be independent and identically distributed (iid) random
variables.
In the modeling of financial time series, conditional
heteroscedasticity is often found, so there is a surge of interest
in the modeling literature [see Baillie et al. (1996), Hauser and
Kunst (1998a,b), Lien and Tse (1999), Elek and M\'{a}rkus (2004), Koopman et al. (2007)]
to extend (\ref{eq:farima}) into the so-called FARIMA-GARCH
 model.
Specifically, for a regular GARCH$(r,s)$ model [cf. Bollerslev (1986)], we
have
\begin{eqnarray}
 \label{eq:garch}
 u_t=\varepsilon_t\sigma_t,~
 \sigma_t^2=\alpha_0+\sum_{i=1}^r\alpha_i u_{t-i}^2+\sum_{i=1}^s\beta_i
 \sigma_{t-i}^2,
\end{eqnarray}
where $\{\varepsilon_t\}$ are iid random variables with zero mean
and unit variance.
 Given a realization $\{X_1,\cdots,X_n\}$ from (\ref{eq:farima})
 with $u_t$ generated by
(\ref{eq:garch}), the joint estimation of the parameter vectors
involved in both FARIMA and GARCH models has been investigated by
Ling and Li (1997).
In practice, one needs to specify the orders of FARIMA and GARCH
models before doing the joint estimation. Hence it's
 customary to estimate the FARIMA model
(\ref{eq:farima}) first, then fit a GARCH model to the residuals
with the orders selected at each stage of model fitting. It is
apparently an  important problem to reassess the
 applicability  of the existing estimator of the parameter vector in
the FARIMA model when $u_t$ is subjected to unknown conditional
heteroscedasticity.

 In this article, we treat the dependence
(including conditional heteroscedasticity) in $\{u_t\}$ nonparametrically.
Specifically, we assume that $\{u_t\}_{t\in\Z}$ is an uncorrelated
mean-zero stationary process  and admits the following representation:
\begin{eqnarray}
\label{eq:ut} u_t=F(\cdots,\varepsilon_{t-1},\varepsilon_t),
\end{eqnarray}
 where $\{\varepsilon_t\}$ are iid random variables and $F$ is a measurable function for which
$u_t$ is a well defined random variable. The framework
(\ref{eq:ut}) is very general and it includes the linear process
$u_t=\sum_{i=0}^{\infty} b_i \varepsilon_{t-i}$ as a special case.
It also includes various nonlinear time series models, such as
bilinear models [Subba Rao and Gabr (1984), Giraitis and Surgalis (2002)], threshold
autoregressive models [Tong (1990)], exponential GARCH [Nelson
(1991)] and asymmetric GARCH models [Ding et al. (1993)].
 One of the major goals of this paper is
to study the asymptotic properties of the Whittle estimator of the
parameter vector involved in (\ref{eq:farima}) when $u_t$ follows
(\ref{eq:ut}).

The framework (\ref{eq:farima}) can be easily extended to allow
nonstationarity. Let
\[ (1-B)^m Y_t=X_t, ~~t=1-m,2-m,\cdots,\]
 where $m\ge 0$ is the
number of times $Y_t$ needs to be differenced to achieve
stationarity. According to Definition 1.1. of Abadir et al. (2007),  $Y_t\sim I(d)$, where $d=d_X+m$.
Alternatively, $Y_t$  is called an $I(d)$ process of type I. Another type of fractional integrated process, that is called
type II process, differs from the Type I counterpart in terms of presample treatment. See Marinucci and Robinson (1999),
Robinson (2005) and Shimotsu and Phillips (2006) for detailed discussions of their differences.
 Estimation of
nonstationary FARIMA processes under parametric assumptions has
been investigated by a few researchers; see Beran (1995), Velasco
and Robinson (2000) and Mayoral (2007) among others. All the work
mentioned above  imposed either conditionally homoscedastic
martingale difference or stronger iid assumptions on $u_t$. Since
the Whittle estimator is not consistent when $d>1$ (see Theorem~\ref{th:inconsistency}),  Velasco and
Robinson (2000) proposed the tapered Whittle estimator and proved  its
consistency and asymptotic normality. Tapering has been frequently
used in the inference of fractionally integrated time series  and
it has nice property of annihilating the nonstationarity. However,
tapering inevitably inflates the variance of the estimator and
therefore results in a loss of efficiency. Recently, in the
context of local Whittle estimation, Abadir et al. (2007)
developed extended Fourier transform and periodogram to handle the
nonstationarity. Here, we generalize their idea to Whittle
estimation and propose the nonstationarity-extended Whittle
estimator, which is shown to be  consistent and asymptotically
normal with higher  efficiency than the tapered Whittle estimator.

The following notation will be used throughout the paper. For a
column vector $x = (x_1, \cdots, x_q)'\in \R^q$, let $|x| =
(\sum_{j=1}^q x_j^2)^{1/2}$. Let $\xi$ be a random vector. Write
$\xi \in {\cal L}^p$ ($p
> 0$) if $\|\xi\|_p := [\E(|\xi|^p )]^{1/p} < \infty$ and let
$\|\cdot\| = \|\cdot\|_2$.  For $\xi \in {\cal L}^1$ define
projection operators ${\cal P}_k \xi = \E( \xi | {\cal F}_k) - \E(
\xi | {\cal F}_{k-1})$, ${\cal F}_k = (\ldots, \varepsilon_{k-1},
\varepsilon_k)$. Let $C>0$, $C_j>0$, $j=1,2,\cdots$ denote generic constants which may
vary from line to line. Denote by $\rightarrow_D$ and
${\rightarrow}_{p}$ convergence in distribution and in
probability, respectively. The symbols $O_{p}(1)$ and $o_{p}(1)$
 signify being bounded in probability and convergence to zero in
probability respectively. Let $N(\mu, \Sigma)$ be a normal
distribution with mean $\mu$ and covariance matrix $\Sigma$. Denote by $\lfloor a\rfloor$  the integer part of $a$.

 The paper is organized as follows. In Section~\ref{sec:m0}, we state
   technical assumptions and derive asymptotic distributional theory for the Whittle estimator  in the stationary case.
 Section~\ref{sec:nonstationary} proves the inconsistency of the Whittle estimator in certain nonstationary region, introduces the nonstationarity-extended Whittle estimator and
discusses its asymptotic properties. In Section~\ref{sec:ss}, we present Monte Carlo simulation results for the Whittle estimator, the tapered Whittle estimator and the nonstationarity-extended Whittle estimator.
Finally, the conclusions are made in Section~\ref{sec:con} and  technical
details are relegated to the Appendix.

\section{Whittle Estimator ($m=0$)}
\label{sec:m0}

Throughout, we consider the following framework, which is more
general than (\ref{eq:farima}).
\begin{eqnarray}
\label{eq:framework} X_t=\sum_{j=0}^{\infty}a_j(\theta)u_{t-j},
~\sum_{j=0}^{\infty}a_j^2(\theta)<\infty, ~a_0(\theta)=1.
\end{eqnarray}
 Let $i=\sqrt{-1}$ be the imaginary unit. For a complex number
 $c$, let $\overline{c}$ be its conjugate.
For a process $\{Z_t\}_{t \in \Z}$, define the periodogram
\begin{eqnarray*}
I_Z(\lambda) = |w_Z(\lambda)|^2, \mbox{ where }
 w_Z(\lambda) = w_{Z,n}(\lambda) = \frac{1}{\sqrt{2\pi n}}
 \sum_{t=1}^n Z_t e^{i t\lambda}.
\end{eqnarray*}
Let $\lambda_j=2\pi j/n ,j=0,1,\cdots,n-1$, be the Fourier
frequencies. Let
$A(\lambda;\theta)=\sum_{j=0}^{\infty}a_j(\theta)e^{ij\lambda}$ be
the transfer function and denote by
$A(\lambda)=A(\lambda;\theta_0)$, where $\theta_0$ is the true
value of $\theta$. Denote by
$G(\lambda;\theta)=|A(\lambda;\theta)|^2$. Then the spectral
density function of $X_t$ is
$f_X(\lambda;\theta)=G(\lambda;\theta)\sigma^2/(2\pi)$, where
$\sigma^2=\var(u_t)$. Denote by
$f_X(\lambda)=f_X(\lambda;\theta_0)$.

  The Whittle estimator $\hat{\theta}_n$ is defined as
\begin{eqnarray}
\hat{\theta}_n=\mbox{argmin}_{\theta\in \Theta}Q_n(\theta),
~~Q_n(\theta)=\frac{2\pi}{n}\sum_{j=1}^{n-1}\frac{I_X(\lambda_j)}{G(\lambda_j;\theta)},
\end{eqnarray}
where $\Theta\subset \R^{s}$ is compact. Further we estimate
$\sigma^2$ by $\hat{\sigma}_n^2=Q_n(\hat{\theta}_n)$. Note that
the zero frequency is excluded in $Q_n(\theta)$ for the purpose of mean correction.

Throughout,  assume that $\theta_0$ lies in the interior of
$\Theta$. In particular,  $d_0=d_{X0}$ is an interior point of
$\Theta^{(1)}=[a_1,a_2]$ with $-1/2<a_1<a_2<\infty$. Hereafter  we use $\theta^{(1)}$ and $\theta^{(-1)}$ to denote the first element and the remaining elements of a vector $\theta$ respectively; $\Theta^{(1)}$ and $\Theta^{(-1)}$ denote the sets for the first element and remaining elements respectively.

To establish the consistency and asymptotic normality of
$\hat{\theta}_n$, we make the following assumptions.

\begin{assumption}
\label{as:longmemo} Assume $f_{X}(\lambda)\sim
|\lambda|^{-2d_{X0}}G$ as $\lambda\rightarrow 0$, where $d_{X0}\in
(-1/2,1/2)$ and $G\in (0,\infty)$. Further we assume that
\[|\partial A(\lambda)/\partial
\lambda|\le C|A(\lambda)||\lambda|^{-1}, ~\lambda\in (0,\pi].\]
\end{assumption}

\begin{assumption}
\label{as:utcum} $\sum_{k_1,k_2,k_3\in \Z} |{\rm
cum}(u_0,u_{k_1},u_{k_2},u_{k_3})|<\infty$.
\end{assumption}
\begin{remark}
\label{rem:utcum} {\rm Summability conditions on joint cumulants are
 widely adopted in spectral analysis. For a linear process
$u_t=\sum_{j\in\Z} b_j \varepsilon_{t-j}$ with $\varepsilon_j$
being iid, Assumption \ref{as:utcum} holds if $\sum_{j\in\Z}
|b_j|<\infty$ and $\varepsilon_1\in {\cal L}^4$. For nonlinear
processes $u_t$, it is satisfied under a geometric moment
contraction (GMC) condition with order 4 [see Wu and Shao's (2004)
Proposition 2]. The process $\{u_t\}$ is GMC with order $\alpha$,
$\alpha > 0$, if there exists a  $\rho=\rho(\alpha)\in (0,1)$ such
that
\begin{eqnarray}
\label{eq:gmc} \E(|u_n^*-u_n|^\alpha)
 \le C\rho^{n},~n\in\N,
\end{eqnarray}
where $u_n^*=F(\cdots, \varepsilon_{-1}', \varepsilon_0',
\varepsilon_1, \cdots, \varepsilon_n)$ and
$\{\varepsilon_t'\}_{t\in \Z}$ is an iid copy of
$\{\varepsilon_t\}_{t\in\Z}$. The property (\ref{eq:gmc})
indicates that the process $\{u_n\}$ forgets its past
exponentially fast and it can be verified for many nonlinear time
series models [Wu and Min (2005), Shao and Wu (2007a)]. Define the
4th cumulant spectral density
\begin{eqnarray*}
f_4(w_1,w_2,w_3)=\frac{1}{(2\pi)^3}
 \sum_{k_1,k_2,k_3\in \Z}\cum(u_0,u_{k_1},u_{k_2},u_{k_3})
 \exp\left(-i\sum_{j=1}^3 w_j k_j\right).
\end{eqnarray*}
Under Assumption \ref{as:utcum}, $f_4(\cdot,\cdot,\cdot)$ is
continuous and bounded. In Shao and Wu (2007b), another set of
sufficient condition for Assumption~\ref{as:utcum} is provided.}
\end{remark}

\begin{assumption}
\label{as:utphys} Suppose $u_t \in {\cal L}^4$. Let $u_k' =
F(\cdots, \varepsilon_{-1}, \varepsilon_0', \varepsilon_1, \cdots,
\varepsilon_k)$ and $\delta_4(k)=\|u_k-u_k'\|_4$.  Assume
$\sum_{k=0}^{\infty}\delta_4(k)<\infty$.
\end{assumption}

\begin{remark}
\label{rem:utphys} {\rm Interpreting (\ref{eq:ut}) as a physical
system, Wu (2005) introduced the physical dependence measure
$\delta_q(k) := \|u_k-u_k'\|_q$, $q\ge 1$.
Intuitively, $\delta_q(\cdot)$
quantifies the dependence of $u_k$ on $\varepsilon_0$ by measuring
the distance between $u_k$ and its coupled version $u_k'$.
Wu (2005) showed that Assumption~\ref{as:utphys} is true
if (\ref{eq:gmc}) holds with $\alpha=4$. In other words, Assumptions~\ref{as:utcum} and ~\ref{as:utphys}
are both implied by the GMC$(4)$ condition, which has been verified for GARCH models of various forms; see Wu and Min (2005) Proposition 3 and Shao and Wu (2007a), Proposition 5.1.
}
\end{remark}

Now we introduce some regularity conditions on
$G(\lambda;\theta)$.
Similar conditions can
be found in Fox and Taqqu (1986), Dahlhaus (1989), Giraitis and
Surgalis (1990) and Velasco and Robinson (2000).

\begin{assumption}
\label{as:whittle}
For any $\delta>0$, the following conditions hold for $\lambda\in
[0,2\pi]$.
\begin{enumerate}
\item The function $\int_{-\pi}^{\pi}\log
G(\lambda;\theta)d\lambda (\equiv 0)$ can be differentiated  twice
under the integral sign.

\item $\theta_1\not=\theta_2$ implies that $\{\lambda:
G(\lambda,\theta_1)\not=G(\lambda,\theta_2)\}$ has positive
Lebesgue measure.

\item $\partial G(\lambda;\theta)/\partial \lambda$ is continuous at all
$(\lambda,\theta)$ except $\lambda=0$, and
\[|G(\lambda;\theta)|\le C|\lambda|^{-2d},~~~\left|\frac{\partial G(\lambda;\theta)}{\partial\lambda}\right|\le C|\lambda|^{-2d-1}. \]

 \item $\partial
G^{-1}(\lambda;\theta)/\partial\theta_j$, $\partial^2
G^{-1}(\lambda;\theta)/\partial\theta_j\partial\theta_k$ and
$\partial^3
G^{-1}(\lambda;\theta)/\partial\theta_j\partial\theta_k\partial\theta_l$ are
continuous at all $(\lambda,\theta)$  except $\lambda=0$, and for $j,k,l=1,\cdots,s$,
\[\left|\frac{\partial G^{-1}(\lambda;\theta)}{\partial \theta_j}\right|\le C|\lambda|^{2d-\delta},\left|\frac{\partial^2 G^{-1}(\lambda;\theta)}{\partial\theta_j\partial\theta_k}\right|\le C|\lambda|^{2d-\delta},~\left|\frac{\partial^3 G^{-1}(\lambda;\theta)}{\partial\theta_j\partial\theta_k\partial\theta_l}\right|\le C|\lambda|^{2d-\delta}. \]

\item $\partial^2G^{-1}(\lambda;\theta)/\partial\lambda
\partial\theta_j$ and $\partial^3G^{-1}(\lambda;\theta)/\partial\lambda
\partial\theta_j\partial\theta_k$  are continuous at all $(\lambda,\theta)$ except
$\lambda=0$, and
\[\left|\frac{\partial^2 G^{-1}(\lambda;\theta)}{\partial \lambda\partial\theta_j}\right|\le C|\lambda|^{2d-1-\delta},~\left|\frac{\partial^3 G^{-1}(\lambda;\theta)}{\partial \lambda\partial\theta_j\partial\theta_k}\right|\le C|\lambda|^{2d-1-\delta},~j,k=1,2,\cdots,s.\]

\end{enumerate}
\end{assumption}

For the FARIMA model defined in (\ref{eq:farima}),
Assumption~\ref{as:whittle} can be easily verified  when $\phi(B)$
and $\psi(B)$ have all roots outside the unit circle.

 Let $W^{(G)}(\theta)$ be the $s\times s$ matrix with $(j,k)$th
entry
\begin{eqnarray*}
W_{jk}^{(G)}(\theta)=\frac{\sigma^2}{2\pi}\int_{0}^{2\pi}G(\lambda;\theta_0)\frac{\partial^2G^{-1}(\lambda;\theta)}{\partial
\theta_j\partial \theta_k}d \lambda
\end{eqnarray*}
and $\Gamma^{(G)}(\theta)$ be the $s\times s$ matrix with
$(j,k)$th entry
\begin{eqnarray*}
\Gamma_{jk}^{(G)}(\theta)&=&\frac{2\sigma^4}{\pi}\int_{0}^{\pi}\frac{\partial\log
G(\lambda;\theta)}{\partial\theta_j}\frac{\partial
\log G(\lambda;\theta)}{\partial\theta_k}d\lambda\\
&&+
8\pi\int_{0}^{\pi}\int_{0}^{\pi}f_4(\lambda_1,-\lambda_2,\lambda_2)\frac{\partial
\log G(\lambda_1;\theta)}{\partial \theta_j}\frac{\partial \log
G(\lambda_2;\theta)}{\partial \theta_k}d\lambda_1 d\lambda_2.
\end{eqnarray*}

\begin{theorem}
\label{th:main} Suppose
Assumptions~\ref{as:longmemo}-\ref{as:whittle} hold. Then
$\hat{\sigma}_n^2\rightarrow_{p}\sigma^2$ and
\begin{eqnarray}
\label{eq:an}
\sqrt{n}(\hat{\theta}_n-\theta_0)\rightarrow_{D}
N(0,W^{(G)}(\theta_0)^{-1}\Gamma^{(G)}(\theta_0)W^{(G)}(\theta_0)^{-1}).
\end{eqnarray}
\end{theorem}
Asymptotic theory for the Whittle estimator has a long history.
Early work by Walker (1964) and Hannan (1973b) dealt with
short-range dependent process. For long-range dependent
  process, see
  Fox and Taqqu (1986), Dahlhaus (1989), Giraitis and Surgailis (1990) and Velasco and Robinson (2000) among others.
    All the works mentioned above assume either Gaussian processes or linear processes with iid or conditionally homescedastic martingale difference
  innovations. In a multivariate setting, Hosoya (1997) obtained the asymptotic normality under certain mixing conditions on the
  conditional moments of $u_t$.
 However, the latter author did not mention how to verify those mixing conditions for statistical
 models. In comparison, our assumptions on $\{u_t\}$ have been verified for
 various nonlinear time series models, including GARCH-type models; see Remarks~\ref{rem:utcum} and
 \ref{rem:utphys}. In general, our Assumption~\ref{as:utphys} is based on physical dependence measure and is not directly comparable to the mixing conditions imposed by Hosoya
 (1997), except in some special cases.  The following example demonstrates that our
 condition is slightly weaker.  On the other hand, we impose the structural assumption (\ref{eq:ut}) on $u_t$ but Hosoya (1997) did not.

 Let
 $u_t=\varepsilon_t\sum_{j=1}^{\infty}a_j\varepsilon_{t-j}$, where
 $\varepsilon_t$ are iid random variables with mean zero, unit
 variance and finite eighth moment. Assume $a_j\sim j^{-\kappa}$.
 Then if $\kappa>1$, our Assumption~\ref{as:utphys} is satisfied.
In the assumption A of Hosoya (1997), it is required that for
$0<t<t_1$,
 \begin{eqnarray}
 \label{eq:hosoya}
 \var(\E(u_{t_1}^2|{\cal F}_t)-\E(u_t^2))=O(|t-t_1|^{-2-\epsilon}),~\mbox{for some}~\epsilon>0.
 \end{eqnarray}
Note that
\begin{eqnarray*}
\mbox{LHS of}~(\ref{eq:hosoya})=\sum_{j=-\infty}^{t}\|{\cal
P}_j\E(u_{t_1}^2|{\cal F}_t)\|^2 =\sum_{j=-\infty}^{t}\|{\cal P}_j
u_{t_1}^2\|^2=\sum_{j=-\infty}^{t}\|{\cal P}_0 u_{t_1-j}^2\|^2.
\end{eqnarray*}
After straightforward calculations, we have ${\cal
P}_0u_{t_1-j}^2=2a_{t_1-j}\varepsilon_0\sum_{k=t_1-j+1}^{\infty}a_{k}\varepsilon_{t_1-j-k}+a_{t_1-j}^2(\varepsilon_0^2-1)$.
So (\ref{eq:hosoya}) holds only when $\kappa>3/2$.

\begin{remark}{\rm
Whittle estimation has been applied to the parametric GARCH models based on squared observations; see Giraitis and Robinson (2001). Note that for the GARCH model (\ref{eq:garch}),
the squared series $\{u_t^2\}$ follows an ARMA($\max(r,s),s$) model [Fan and Yao (2003)], i.e.
\[u_t^2=\alpha_0+\sum_{i=1}^{\max(r,s)}(\alpha_i+\beta_i)u_{t-i}^2+e_t-\sum_{j=1}^{s} \beta_j e_{t-j},\]
where $\alpha_{r+j}=\beta_{s+j}=0$ for $j\ge 1$, $e_t=X_t^2-\sigma_t^2$
is a martingale difference sequence. Giraitis and Robinson (2001) adopted a  more general framework and obtained a central limit theorem for the Whittle estimator  under an $8$-th moment condition on $u_t$. Note that Ling and McAleer (2002) provided a sufficient and necessary condition for the existence of the eighth moment for $u_t$, which implies that $u_t=G(\cdots,\varepsilon_{t-1},\varepsilon_{t})$ for some measurable function $G$ and $u_t$ is GMC with order $8$; see  Proposition 5.1 in Shao and Wu (2007a). Following the argument in the latter paper,  it is not hard to show that $e_t$ admits a nonlinear causal representation, i.e. $e_t=J(\cdots,\varepsilon_{t-1},\varepsilon_{t})$ for some measurable function $J$, and $e_t$ is GMC of order $4$. In view of Remarks~\ref{rem:utcum} and \ref{rem:utphys},  our Assumptions~\ref{as:utcum} and \ref{as:utphys} are implied by GMC(4), so our Theorem~\ref{th:main} is directly applicable to this setting.

It is worth mentioning  the work of
Zaffaroni and d'Italia (2003), who studied Whittle estimation of
long memory volatility models with ARMA levels. It seems our result is not applicable to that setting.
In addition, there are models that allow long memory  in both conditional mean and conditional variance  [cf. Giraitis and Surgalis (2002)], for which our theory no longer applies. Under our framework, we allow long memory in the level but the square of the conditional heteroscedastic error needs to be short-range dependent, so we exclude models that have long memory in volatility.

}
\end{remark}

\begin{remark}
{\rm  The asymptotic covariance matrix in (\ref{eq:an}) admits a different form compared to those in the literature. Below we show that under extra conditions
 on $u_t$, our asymptotic covariance matrix is the same as those existing ones; compare Fox and Taqqu (1986), Dahlhaus (1989), Giraitis and Surgalis (1990) and Velasco and Robinson (2000). A key assumption in  Velasco and Robinson (2000) is that
\begin{eqnarray}
\label{eq:mds}
\E(u_t|{\cal F}_{t-1})=0, ~\E(u_t^j|{\cal F}_{t-1})=\mbox{constant}~\mbox{for}~ j=2,3,4.
\end{eqnarray}
 Under (\ref{eq:mds}), $\cum(u_0,u_{k_1},u_{k_2},u_{k_3})=\E(u_0^4)-3\sigma^4$ only when $k_1=k_2=k_3=0$ and zero otherwise. This implies that $f_4(w_1,w_2,w_3)=[\E(u_0^4)-3\sigma^4]/(2\pi)^3$ for all $(w_1,w_2,w_3)\in [-\pi,\pi)^3$. If $u_t$ are iid or Gaussian [see Fox and Taqqu (1986), Dahlhaus (1989) and Giraitis and Surgalis (1990)], the fourth order spectrum is also a constant. Consequently,
 \[\Gamma_{jk}^{(G)}(\theta)=\frac{2\sigma^4}{\pi} \int_{0}^{\pi}\frac{\partial\log
G(\lambda;\theta)}{\partial\theta_j}\frac{\partial
\log G(\lambda;\theta)}{\partial\theta_k}d\lambda\]
and the asymptotic covariance matrix in (\ref{eq:an}) reduces to $4\pi \Sigma(\theta_0)^{-1}$, where the $(j,k)$th entry for $\Sigma(\theta_0)$ is
\[\int_{-\pi}^{\pi}\frac{\partial\log
G(\lambda;\theta_0)}{\partial\theta_j}\frac{\partial
\log G(\lambda;\theta_0)}{\partial\theta_k}d\lambda.\]
Note that we have applied the fact that $W_{jk}^{(G)}(\theta_0)=(\sigma^2/\pi)\int_0^{\pi}\frac{\partial\log G(\lambda;\theta_0)}{\partial\theta_j}\frac{\partial \log G(\lambda;\theta_0)}{\partial\theta_k}d\lambda
$.
Hence, our asymptotic covariance matrix coincides with those presented in Theorem 4 of Giraitis and Surgailis (1990) and Theorem 2 of Velasco and Robinson (2000) for $p=1$, i.e. the untapered case. Theorem~\ref{th:main} suggests that conditional heteroscedasticity affects  the asymptotic covariance matrix through the non-constant fourth order cumulant spectra of $u_t$.
   For some non-Gaussian
processes,  Giraitis and Taqqu (1999) demonstrated that the Whittle
estimator may not be $\sqrt{n}$-consistent and the limiting
distribution may not be Gaussian.   Our results have
different applicabilities.

To construct a confidence region for $\theta_0$, one can  estimate the asymptotic covariance matrix directly, which involves the estimation of the integral of the fourth order cumulant spectra. For short memory time series, the latter problem has
been studied by Taniguchi (1982), Keenan (1987) and Chiu (1988), but the applicability of their methods to long memory time series is not clear.
Alternative methods that bypass direction estimation are currently under investigation and we hope to report that in the near future.

}
\end{remark}

\begin{remark}
{\rm
Assumption~\ref{as:longmemo} excludes seasonal long memory processes, such as Gegenbauer process [Gray, Zhang and Woodward (1989)], in which the spectral density function has
 a pole at a nonzero frequency. The work by Velasco and Robinson (2000) seems to allow for such processes.
}
\end{remark}

\section{Nonstationary case}
\label{sec:nonstationary}

In this section, we shall consider the nonstationary case, i.e. $m\ge 1$. For the convenience of presentation, we assume that
$G(\lambda;\theta)=|1-e^{i\lambda}|^{-2d_X} \tilde{G}(\lambda;\theta^{(-1)})$, i.e. the spectral density function of $X_t$ can be factorized into
a product of the fractional integrated component $|1-e^{i\lambda}|^{-2d_X}$ and the short memory component $\tilde{G}(\lambda;\theta^{(-1)})$.  This adds a slight constraint for the class of models, but is not overly restrictive due to the prevalence of the fractionally integrated models  in practice. Define $H(\lambda;\theta)=|1-e^{i\lambda}|^{-2d}\tilde{G}(\lambda;\theta^{(-1)})$, where $d=m+d_X$ is the fractional integration order of $Y_t$. Note that $m=\lfloor d+1/2\rfloor$.

\subsection{Inconsistency of the Whittle estimator when $d_0>1$}

The consistency  of the Whittle estimator has been obtained by Velasco and Robinson (2000) for $d_0\in (1/2,1)$ and it is still unknown whether the Whittle estimator is consistent when $d_0\ge 1$. A semiparametric frequency-domain approach to estimating the order of fractional integration, that is closely related to the Whittle estimation, is the so-called local Whittle estimation.
The local Whittle estimator of $d$ is consistent when $d_0\in (-1/2,1/2)\cup (1/2,1]$ and is inconsistent when $d_0>1$; see Phillips and Shimotsu (2004) and Shao and Wu (2007b).
 Similar results can be expected for the Whittle estimator due to the similarity in the theoretical justifications for these two estimators.  Here we shall show the inconsistency of the Whittle estimator when $d_0\in (1,3/2)$, which provides a sound motivation for the consideration of the nonstationarity-extended Whittle estimation (see Section~\ref{sec:m1}). Using a similar argument, one can show that when $d_0\in \{d>3/2,d\not=(2k+1)/2,k=2,3,\cdots\}$, the Whittle estimator is inconsistent. Since the proof does not involve additional methodological difficulties, we omit the details.

Define
\[{\theta}_n^*=\mbox{argmin}_{\theta\in\Theta} M_n(\theta),~ M_n(\theta):=\frac{2\pi}{n}\sum_{j=1}^{n-1}\frac{I_Y(\lambda_j)}{H(\lambda_j;\theta)}\]
Denote by $\theta^*=\mbox{argmin}_{\theta\in\Theta} M(\theta)$, where $M(\theta) = \int_0^{2\pi} |1-e^{i\lambda}|^{-2}H^{-1}(\lambda;\theta) d\lambda$.
Let $K(\lambda;\theta)=|1-e^{i\lambda}|^{2d}H(\lambda;\theta)$.
\begin{theorem}
\label{th:inconsistency}
Suppose Assumptions~\ref{as:longmemo}-\ref{as:utphys} and Assumption~\ref{as:whittle} (with $G(\lambda;\theta)$ replaced by $H(\lambda;\theta)$) hold.  Further, for all
 $\lambda$ and $\theta$, $0<C_1<K(\lambda;\theta)<C_2<\infty$. Assuming $d_0\in (1,3/2)$, then
  ${\theta}_n^*\rightarrow_p \theta^*$.
\end{theorem}

  Like the local Whittle estimator,  the Whittle estimator of $d$ converges to $1$ in probability when $d_0\in (1,3/2)$, since $\theta^{*(1)}=1$. Further, we conjecture that when $d_0=1$, the Whittle estimator is also consistent.

\subsection{Nonstationarity-extended Whittle Estimator}
\label{sec:m1}

In this subsection, we propose the
 nonstationarity-extended Whittle estimator following the idea of
 Abadir et al. (2007),  who introduced the extended Fourier transform and
periodogram to deal with nonstationarity in the local Whittle
estimation. Note that the  extended discrete Fourier transform and periodogram have been suggested in an early work by
Phillips (1999) for $d_0\in (-1/2,3/2)$.

 Assume $d_0=m_0+d_{X0}\notin \{p-1/2,p\in\Z\}$, where $m_0$ is the true value of $m$.
 We define the extended periodogram as
\begin{eqnarray*}
I_Y(\lambda_j;d):=|w_Y(\lambda_j;d)|^2,
~~~w_Y(\lambda_j;d)=w_Y(\lambda_j)+J(\lambda_j;d),~d\in [a_1,a_2],
\end{eqnarray*}
where
\begin{eqnarray*}
J(\lambda_j;d)=\left\{\begin{array}{cc} 0&\mbox{if}~d\in
[-1/2,1/2),\\
e^{i\lambda_j}\sum_{r=1}^{m}(1-e^{i\lambda_j})^{-r}Z_r,&\mbox{if}~d\in
I_m:= [m-1/2,m+1/2), m\in\N,
\end{array}\right.
\end{eqnarray*}
with $Z_r=(2\pi n)^{-1/2}((1-B)^{r-1}Y_n-(1-B)^{r-1}Y_0),
r=1,2,\cdots,m$. As mentioned in Abadir et al. (2007), the enumeration of the data should be $Y_{-h+1},\cdots,Y_n$, where $h=\lfloor a_2+1/2\rfloor \vee 0$.  For example, when  $a_2<1/2$, the enumeration is $Y_1,\cdots,Y_n$; when $a_2\in [1/2,3/2)$, the data is enumerated as $Y_0,\cdots,Y_n$.

Then the nonstationarity-extended Whittle estimator is defined as
\begin{eqnarray*}
\check{\theta}_n=\mbox{argmin}_{\theta\in\Theta}L_n(\theta),~~L_n(\theta)=\frac{2\pi}{n}\sum_{j=1}^{n-1}\frac{I_Y(\lambda_j;d)}{H(\lambda_j;\theta)}.
\end{eqnarray*}

Again, $\sigma^2$ is estimated by
$\check{\sigma}_n^2=L_n(\check{\theta}_n)$.

\begin{theorem}
\label{th:consistency}
 Suppose Assumptions~\ref{as:longmemo}-\ref{as:utphys} and Assumption~\ref{as:whittle} (with $G(\lambda;\theta)$ replaced by $H(\lambda;\theta)$) hold.  Further, for all
 $\lambda$ and $\theta$, $0<C_1<K(\lambda;\theta)<C_2<\infty$.
Then  we have
  $\check{\sigma}_n^2\rightarrow_{p} \sigma^2$ and
 \begin{eqnarray*}
 \sqrt{n}(\check{\theta}_n-\theta_0)\rightarrow_{D} N(0,W^{(H)}(\theta_0)^{-1}\Gamma^{(H)}(\theta_0)W^{(H)}(\theta_0)^{-1}).
 \end{eqnarray*}
\end{theorem}

Therefore, the nonstationarity-extended Whittle estimator is consistent and
asymptotically normal irrespective of the true value of the fractional integration order. Further, the asymptotic covariance matrix admits the same form as the stationary case, whereas for the tapered Whittle estimator [Velasco and Robinson (2000)], the variance is inflated due to the exclusion  of certain frequencies and the tapering effect, and  the inflation factor gets large   as the order of fractional integration increases since a higher order taper is needed to accommodate a larger $d$.

\section{Finite Sample Performance}
\label{sec:ss}

In this section, we examine the finite sample performance of the Whittle estimator $\hat{\theta}_n$ ($\theta_n^*$), the nonstationarity-extended Whittle estimator $\check{\theta}_n$, and two tapered Whittle estimators proposed by Velasco and Robinson (2000) through a small simulation study. For tapered Whittle estimators, we use both cosine weights, i.e. $h_t=(1-\cos(2\pi t/n))/2$  and Parzen's weights, where
\begin{eqnarray*}
h_t=\left\{
\begin{array}{cc}
1-6[|\frac{2t-n}{n}|^2-|\frac{2t-n}{n}|^3],&N<t<3N\\
2\{1-|\frac{2t-n}{n}|\}^3,&1\le t\le N, ~3N\le t\le 4N
\end{array}
\right.
\end{eqnarray*}
with $N=n/4$. So the number of frequencies included in the objective functions of tapered Whittle estimators
 are $\lfloor n/3-1\rfloor$ and $\lfloor n/4-1\rfloor$ respectively.
Two sample sizes $n=200$ and $n=512$ are investigated.

 Consider the following model
\begin{eqnarray}
\label{eq:farima2}
(1-0.65 B+0.6 B^2)(1-B)^d Y_t=u_t,
\end{eqnarray}
where
\begin{eqnarray}
\label{eq:garch4}
u_t=\varepsilon_t\sigma_t,~\sigma_t^2=0.4+0.3u_{t-1}^2+0.3\sigma_{t-1}^2
\end{eqnarray}
 with $ \varepsilon_t$ independently generated from  standard normal distribution. Thus the GARCH model (\ref{eq:garch4}) admits a finite fourth moment; see Davidson (2004) for some sufficient conditions on the existence of higher order moments for GARCH models.
  Note that the FARIMA(2,d,0) model (\ref{eq:farima2}) has been investigated in Velasco and Robinson (2000), but under iid assumptions on $u_t$. Here we examine a wide range of $d$'s from $-0.4$ to $2.4$, including $d=1.5$, which is not covered by our theory.  We take $[a_1,a_2]=[-.49,3.49]$. Tables~\ref{tb:table1}-\ref{tb:table3} report the bias and $100\times$mean square error (MSE) for the estimates of $d$, $\phi_1$ and $\phi_2$   based on 1000 replications.  In the tables, the symbols W-1, W-3, W-4 and EW correspond to the Whittle estimator, tapered Whittle estimator with cosine weights, tapered Whittle estimator with Parzen's weights,  and the nonstationarity-extended Whittle estimator respectively.

As we expected, the bias and MSE decrease as the sample size increases. It appears that the mean squared error for the nonstationarity-extended Whittle estimator is substantially smaller than those for two tapered estimators, and is similar to that for the Whittle estimator when $d_0\in (-0.5,0.5)$. The inconsistency of the Whittle estimator when $d_0>1$ can be easily seen from both the bias and MSE. The tapered estimator using Parzen's weights shows a severe downward bias in estimating $d$, upward bias in estimating $\phi_1$ when $n=200$. Although in theory, the tapered estimator with cosine weights is still asymptotically normal when $d_0=2.4$, the bias and MSE get noticeably large since it is close to the region of inconsistency, i.e. $d_0>2.5$. The result for the case $d_0=1.5$ does not seem to be very much different from those for other $d$s, which suggests the theory works for this case.

We also tried three different models for $u_t$: (i) iid N(0,1); (ii) asymmetric GARCH(1,1); (iii) regular GARCH(1,1) but with infinite fourth moments.  The results are qualitatively similar to what we observe here (results not shown). Overall, the nonstationarity-extended Whittle estimator  outperforms both tapered Whittle estimators uniformly in the range of $d$ examined here.
Both theory and simulation studies suggest that the nonstationarity-extended Whittle estimator is preferable to the tapered Whittle estimator, so we recommend its use to the practitioners.

\section{Conclusions}
\label{sec:con}

This paper presents an asymptotic theory for the Whittle estimator of a class of long memory time series
models with uncorrelated but dependent errors. Our dependence conditions on the errors are mild and can be verified for a large class of nonlinear time series models, including GARCH-type models. Following the idea in Abadir et al. (2007), we extend the range of consistency
and asymptotic normality by developing the nonstationarity-extended Whittle estimator. Both theory and finite sample results demonstrate that
the proposed estimator is more efficient than the tapered Whittle estimator [Velasco and Robinson (2000)]. It is worth noting that our framework is limited to Type I fractional process.
For Type II process, the extended local Whittle estimation has been investigated by Shimotsu and Phillips (2005) and it would be  interesting to extend their idea to Whittle estimation.

\section{Technical Appendix}

 For the convenience of notation, write $A_j=A(\lambda_j)$,
$I_{Xj}=I_X(\lambda_j)$, $I_{uj}=I_u(\lambda_j)$,
$f_{Xj}=f_X(\lambda_j)$, $f_{uj}=f_{u}(\lambda_j)$,
$\tilde{I}_{Xj}=I_{Xj}f_{Xj}^{-1}$,
$\tilde{I}_{uj}=I_{uj}f_{uj}^{-1}$, $w_{Xj}=w_X(\lambda_j)$,
$w_{uj}= w_u(\lambda_j)$,
$j=1,\cdots, n-1$. Let
$g_j=w_{Xj}/\sqrt{f_{Xj}}$ and $h_j=w_{uj}/\sqrt{f_{uj}}$. Denote
by $D(w) = D_n(w)=\sum_{t=1}^n e^{i t w}$. Let $K(w)=(2\pi
n)^{-1}|D(w)|^2$ be Fej\'er's kernel. Denote by $a\vee
b=\max(a,b)$ and $a\wedge b=\min(a,b)$. Let $\tilde{n}:=\lfloor n/2\rfloor$.

 \subsection{Proofs of Theorems~\ref{th:main}, \ref{th:inconsistency} and~\ref{th:consistency}}

\noindent Proof of Theorem~\ref{th:main}: The consistency of
$\hat{\theta}_n$ can be proved along the line in the proof of
Theorem 1 of Velasco and Robinson (2000). Since it is simpler than
the proof of the consistency of $\check{\theta}_n$ (see
Theorem~\ref{th:consistency}), we skip the presentation.

Applying the mean value theorem, we have
\[0=\frac{\partial Q_n(\hat{\theta}_n)}{\partial \theta}=\frac{\partial Q_n(\theta_0)}{\partial \theta}+\frac{\partial^2Q_n(\breve{\theta}_n)}{\partial\theta^2}(\hat{\theta}_n-\theta_0),\]
where $\breve{\theta}_n=\theta_0+\alpha(\hat{\theta}_n-\theta_0)$
for some $\alpha\in (0,1)$. Under Assumption~\ref{as:whittle}, we
get by Lemma~\ref{lem:uniform} that
$\partial^2Q_n(\theta)/\partial\theta^2 \rightarrow_{p}
W^{(G)}(\theta)$ elementwise uniformly in $\theta\in\Theta_1$,
where $\Theta_1=[d_0-1/2+\Delta,a_2]\times \Theta^{(-1)}$ for some $\Delta\in
(0,1/4)$. Following the same argument as in the proof of Velasco
and Robinson's Lemma 7 (2000),
$\sqrt{n}(\hat{\theta}_n-\theta_0)=W^{(G)}(\theta_0)^{-1}\sqrt{n}\partial
Q_n(\theta_0)/\partial \theta+o_p(1)$.
 Thus the conclusion follows if we can show
\begin{eqnarray}
\sqrt{n}\frac{\partial Q_n(\theta_0)}{\partial
\theta}=-\frac{\sigma^2}{\sqrt{n}}\sum_{j=1}^{n-1}\tilde{I}_{Xj}\frac{\partial
\log G(\lambda_j;\theta_0)}{\partial\theta} \rightarrow_{D}
N(0,\Gamma^{(G)}(\theta_0)).
\end{eqnarray}
For each   $h=1,2,\cdots,s$, let $l(\lambda)=\partial\log
G(\lambda;\theta_0)/\partial\theta_h$ and $l_k=l(\lambda_k)$. Note that $l(\lambda_k)=-G(\lambda_k;\theta_0)\partial G^{-1}(\lambda_k;\theta_0)/\partial \theta_h$.
Under Assumption~\ref{as:whittle}, we apply the mean value theorem and obtain   $|l_k-l_{k+1}|\le
Cn^{-1}\lambda_k^{-1-\delta}$ via a straightforward calculation. By Lemma~\ref{lem:bartlet3},
\begin{eqnarray*}
&&\hspace{-0.8cm}\E\left|\sum_{k=1}^{\tilde{n}}l_k(\tilde{I}_{Xk}-\tilde{I}_{uk})\right|
\le\sum_{k=1}^{\tilde{n}}|l_k-l_{k+1}|\E\left|\sum_{j=1}^{k}(\tilde{I}_{Xj}-\tilde{I}_{uj})\right|+|l_{\tilde{n}}-l_{\tilde{n}+1}|\E\left|\sum_{j=1}^{\tilde{n}}(\tilde{I}_{Xj}-\tilde{I}_{uj})\right|\\
&&\hspace{0.5cm}\le
\frac{C}{n}\sum_{k=1}^{\tilde{n}}\lambda_k^{-1-\delta}(k^{1/4}(1+\log
k)^{1/2}+k^{1/2}n^{-1/4})+Cn^{1/4}\log n=o(\sqrt{n}).
\end{eqnarray*}
Thus it suffices to show
\begin{eqnarray*}
\frac{2\pi}{\sqrt{n}}\sum_{j=1}^{n-1}I_{uj}\frac{\partial\log
G(\lambda_j;\theta_0)}{\partial\theta} \rightarrow_{D}
N(0,\Gamma^{(G)}(\theta_0)),
\end{eqnarray*}
which is established in Lemma~\ref{lem:cltu}. Finally, the
consistency of $\hat{\sigma}_n^2$ follows from the consistency of
$\hat{\theta}_n$ and Lemma~\ref{lem:uniform}. The proof is now
complete. \qed

\bigskip

\noindent Proof of Theorem~\ref{th:inconsistency}:  Let $\Theta_1=[1/2+\Delta,a_2]\times \Theta^{(-1)}$ and $\Theta_2=[a_1,1/2+\Delta)\times \Theta^{(-1)}$, possibly empty. When
$d_0\in (1,3/2)$, $m_0=1$ and $d_{X0}\in (0,1/2)$. Since $w_Y(\lambda_j)=w_Y(\lambda_j;d_0)-J(\lambda_j;d_0)$ and $w_Y(\lambda_j;d_0)=w_X(\lambda_j)(1-e^{i\lambda_j})^{-1}$, we can write
$M_n(\theta)=M_{1n}(\theta)-M_{2n}(\theta)-\overline{M_{2n}(\theta)}+M_{3n}(\theta)$, where
\begin{eqnarray*}
M_{1n}(\theta)&=&\frac{2\pi}{n}\sum_{j=1}^{n-1}\frac{I_X(\lambda_j)|1-e^{i\lambda_j}|^{-2}}{H(\lambda_j;\theta)},~M_{2n}(\theta)=\frac{2\pi}{n}\sum_{j=1}^{n-1}\frac{w_Y(\lambda_j;d_0)\overline{J(\lambda_j;d_0)}}{H(\lambda_j;\theta)}\\
M_{3n}(\theta)&=&\frac{2\pi}{n}\sum_{j=1}^{n-1}\frac{|J(\lambda_j;d_0)|^2}{H(\lambda_j;\theta)}=\frac{2\pi}{n}\sum_{j=1}^{n-1}\frac{|1-e^{i\lambda_j}|^{-2}}{H(\lambda_j;\theta)}Z_1^2,
\end{eqnarray*}
with $Z_1=(2\pi n)^{-1/2}\sum_{t=1}^{n}X_t$. Applying the argument in Lemma A.5 of Shao and Wu (2007b), it is not hard to show that
\begin{eqnarray}
\label{eq:Z1}
0<C_3<\lim_{n\rightarrow\infty}\E(Z_1^2)/n^{2d_0-2}<C_4<\infty.
\end{eqnarray}
Hence for $\theta\in \Theta_1$, $M_{3n}(\theta)$ dominates the other two terms in magnitude.  By Lemma~\ref{lem:uniform}, $\sup_{\theta\in\Theta_1} |M_{1n}(\theta)-\bar{M}(\theta)|=o_p(1)$, where $\bar{M}(\theta)=\int_{-\pi}^{\pi}f_X(\lambda)|1-e^{i\lambda}|^{-2}H^{-1}(\lambda;\theta)d\lambda$. Under Assumption~\ref{as:whittle}, we have that $M_{3n}(\theta)/Z_1^2\rightarrow M(\theta)$ holds uniformly for $\theta\in\Theta_1$. By the Cauchy-Schwarz inequality, it follows that
$\sup_{\theta\in\Theta_1}|M_n(\theta)/Z_1^2-M(\theta)|\rightarrow_p 0$ and  that  with probability tending to $1$,
\[\inf_{|\theta-\theta^*|\ge \epsilon,~\theta\in\Theta_1}\{M_n(\theta)/Z_1^2-M_n(\theta^*)/Z_1^2\}\ge \eta(\epsilon)>0, ~\mbox{for any}~\epsilon>0,\]
since $M(\theta)$ is uniformly  continuous in $\Theta_1$. In view of the argument in Velasco and Robinson (2000), the conclusion follows if we can show that for any $\epsilon>0$,
\begin{eqnarray}
\label{eq:anothercond}
P\left(\inf_{\theta\in\Theta_2}\{M_n(\theta)/Z_1^2-M(\theta^*)\}\le \epsilon\right)\rightarrow 0.
\end{eqnarray}

  Denote by
$p_n=\lfloor n/3\rfloor$, $a_j=(j/p_n)^{2\Delta-1}$ if $1\le j\le p_n$. For $\theta\in \Theta_2$,
\begin{eqnarray*}
M_n(\theta)&\ge& \frac{2\pi }{nC_2}\sum_{j=1}^{n-1}|1-e^{i\lambda_j}|^{2d}I_Y(\lambda_j)\ge  \frac{C}{n} \sum_{j=1}^{p_n}\lambda_j^{2d}I_Y(\lambda_j)\\
&\ge&\frac{C}{n^3}\sum_{j=1}^{p_n}(j/p_n)^{2d-2}j^2 I_Y(\lambda_j)\ge \frac{C}{n^3}\sum_{j=1}^{p_n}a_j j^2 I_Y(\lambda_j).
\end{eqnarray*}
Again, by Lemma~\ref{lem:uniform}, the Cauchy-Schwarz inequality and (\ref{eq:Z1}), we have
\begin{eqnarray*}
\inf_{\theta\in\Theta_2} M_n(\theta)/Z_1^2\ge \frac{C}{n^3}\sum_{j=1}^{p_n}a_j j^2|1-e^{i\lambda_j}|^{-2}(1+o_p(1))\ge \frac{C}{n}\sum_{j=1}^{p_n}a_j(1+o_p(1)),
\end{eqnarray*}
where the above constant $C$ does not depend on $\Delta$. Since
\[n^{-1}\sum_{j=1}^{p_n}a_j\rightarrow \frac{1}{3} \int_0^1 x^{2\Delta-1}dx=(6\Delta)^{-1},\]
 which can be made arbitrary large when $\Delta>0$ is sufficiently small. Thus (\ref{eq:anothercond}) holds since $M(\theta^*)$ is finite. This completes the proof.

\qed

\bigskip

\noindent Proof of Theorem~\ref{th:consistency}:
The proof of the consistency closely follows the argument in the
proof of Velasco and Robinson's (2000) Theorem 1.
By definition, we have
\[I_Y(\lambda_j;d)=|w_{Y}(\lambda_j;d_0)+\tau_j(d)|^2, ~\mbox{where}~\tau_j(d)=J(\lambda_j;d)-J(\lambda_j;d_0), ~d\in [a_1,a_2].\]
So  $\tau_j(d)=0$ if $m=m_0$. By Lemma 4.4 in Abadir et al.
(2007),
$w_{Y}(\lambda_j;d_0)=w_{X}(\lambda_j)(1-e^{i\lambda_j})^{-m_0}$.
Write
$L_n(\theta)=L_{1n}(\theta)+L_{2n}(\theta)+\overline{L_{2n}(\theta)}+L_{3n}(\theta)$,
where
\begin{eqnarray}
\label{eq:L123n}
L_{1n}(\theta)&=&\frac{2\pi}{n}\sum_{j=1}^{n-1}\frac{|w_{Y}(\lambda_j;d_0)|^2}{H(\lambda_j;\theta)}=\frac{2\pi}{n}\sum_{j=1}^{n-1}\frac{I_X(\lambda_j)|1-e^{i\lambda_j}|^{-2m_0}}{H(\lambda_j;\theta)}\nonumber\\
L_{2n}(\theta)&=&\frac{2\pi}{n}\sum_{j=1}^{n-1}\frac{w_Y(\lambda_j;d_0)\overline{\tau_j(d)}}{H(\lambda_j;\theta)},~~L_{3n}(\theta)=\frac{2\pi}{n}\sum_{j=1}^{n-1}\frac{|\tau_j(d)|^2}{H(\lambda_j;\theta)}.
\end{eqnarray}
Define
\begin{eqnarray}
\label{eq:checkln} \check{L}_{1n}(\theta)=\frac{2\pi}{
n}\sum_{j=1}^{n-1}\frac{f_X(\lambda_j)|1-e^{i\lambda_j}|^{-2m_0}}{H(\lambda_j;\theta)},~
L(\theta)=\int_{-\pi}^{\pi}\frac{f_{X}(\lambda)|1-e^{i\lambda}|^{-2m_0}}{H(\lambda;\theta)}d\lambda.
\end{eqnarray}
 For any $\Delta\in (0,1/4)$, let $d_{\Delta}=d_0-1/2+\Delta$ and define
$\Theta_1=[d_{\Delta},a_2]\times \Theta^{(-1)}$ and
$\Theta_2=[a_1,d_{\Delta})\times \Theta^{(-1)}$, possibly empty.
In view of the argument in Velasco and Robinson (2000), it
suffices to show that as $n\rightarrow\infty$, with probability
tending to 1,
\begin{eqnarray}
\label{eq:1st}
 \inf_{|\theta-\theta_0|\ge
\epsilon,\theta\in\Theta_1}\{{L}_n(\theta)-{L}_n(\theta_0)\}\ge
\eta(\epsilon)>0 ~\mbox{for any}~ \epsilon>0
\end{eqnarray}
and
\begin{eqnarray}
\label{eq:2nd}
P\left(\inf_{\theta\in\Theta_2}\{{L}_n(\theta)-{L}(\theta_0)\}\le
\epsilon\right)\rightarrow 0~\mbox{for any}~\epsilon>0.
\end{eqnarray}
The statement (\ref{eq:1st}) is implied by (i)
$\inf_{|\theta-\theta_0|>\epsilon,
\theta\in\Theta_1}|L(\theta)-L(\theta_0)|\ge \eta(\epsilon)>0$ and
(ii).
$\sup_{\theta\in\Theta_1}|{L}_n(\theta)-L(\theta)|\rightarrow_{p}
0$. The former follows from the uniform continuity of $L(\theta)$
on $\Theta_1$ and the identifiability conditions in
Assumption~\ref{as:whittle}. It follows from
Lemma~\ref{lem:uniform} that
$\sup_{\theta\in\Theta_1}|L_{1n}(\theta)-L(\theta)|=o_p(1)$, which
consequently results in (ii) in view of Lemma~\ref{lem:Ln} and the
fact that
$\sup_{\theta\in\Theta_1}|\check{L}_{1n}(\theta)-L(\theta)|=o(1)$.

Next, we  show (\ref{eq:2nd}) when $\Theta_2$ is nonempty. By
Lemma~\ref{lem:lower}, we have
that with probability tending to $1$,
\begin{eqnarray*}
\inf_{\theta\in\Theta_2}
{L}_n(\theta)
&\ge&\frac{C}{n}\sum_{j=1}^{\tilde{n}}\lambda_j^{2d_{\Delta}}I_{Xj}|1-e^{i\lambda_j}|^{-2m_0}
\end{eqnarray*}
where the positive constant $C$ above is independent of $\Delta$.
By Lemma~\ref{lem:uniform}, the above term converges in
probability to
\begin{eqnarray*}
\frac{
C}{n}\sum_{j=1}^{\tilde{n}}f_X(\lambda_j)|1-e^{i\lambda_j}|^{-2m_0}|\lambda_j|^{2d_{\Delta}}&\ge&
\frac{C}{n}\sum_{j=1}^{\tilde{n}}|\lambda_j|^{2\Delta-1}\sim
C\int_{0}^{\pi}|\lambda|^{2\Delta-1}d\lambda\\
&=&C\pi^{2\Delta}/\Delta\rightarrow\infty~\mbox{as}~\Delta\downarrow
0.
\end{eqnarray*}
So the assertion (\ref{eq:2nd}) follows by choosing $\Delta>0$
such that $C\pi^{2\Delta}/\Delta>L(\theta_0)+2\epsilon$.
Therefore, $\check{\theta}_n\rightarrow_{p}\theta_0$.

To show the asymptotic normality of $\check{\theta}_n$, we define
another (infeasible) estimator $\tilde{\theta}_n$ by
\[\tilde{\theta}_n=\mbox{argmin}_{\theta\in\Theta}L_{1n}(\theta).\]
Since $\check{\theta}_n^{(1)}\rightarrow_{p}d_0$ and
$d_0\not=p+1/2$, $p\in\Z$, $P(\check{\theta}_n^{(1)}\in
I_{m_0})\rightarrow 1$. So
$P(\check{\theta}_n\not=\tilde{\theta}_n)\rightarrow 0$ and
$\tilde{\theta}_n\rightarrow_p\theta_0$. Thus it suffices to show
the asymptotic normality of $\tilde{\theta}_n$.
Note that
\begin{eqnarray*}
\sqrt{n}\frac{\partial L_{1n}(\theta_0)}{\partial
\theta}
&=&\frac{2\pi}{\sqrt{n}}\sum_{j=1}^{n-1}\frac{I_X(\lambda_j)}{|1-e^{i\lambda_j}|^{2m_0}}\frac{\partial
H^{-1}(\lambda_j;\theta_0)}{\partial\theta}\\
&=&-\frac{2\pi}{\sqrt{n}}\sum_{j=1}^{n-1}\frac{I_X(\lambda_j)}{G(\lambda_j;\theta_0)}\frac{\partial\log
H(\lambda_j;\theta_0)}{\partial\theta}.
\end{eqnarray*}
By a similar argument as in the proof of Theorem~\ref{th:main} and
Lemma~\ref{lem:cltu}, we can show that
\begin{eqnarray*}
\sqrt{n}\frac{\partial L_{1n}(\theta_0)}{\partial
\theta}&=&-\frac{2\pi}{\sqrt{n}}\sum_{j=1}^{n-1}
I_{u}(\lambda_j)\frac{\partial\log
H(\lambda_j;\theta_0)}{\partial\theta}+o_p(1)\rightarrow_{D}
N(0,\Gamma^{(H)}(\theta_0)).
\end{eqnarray*}
Further, by Lemma~\ref{lem:uniform}, $\partial^2
L_{1n}(\theta)/\partial^2\theta\rightarrow_p W^{(H)}(\theta)$
uniformly in $\theta\in\Theta_1$ elementwise. Thus the asymptotic
normality of $\tilde{\theta}_n$ holds and so does
$\check{\theta}_n$.
Finally, it follows from  the consistency of
$\check{\theta}_n$, the continuity of $H^{-1}(\lambda;\theta)$
with respect to $\theta$ and Lemma~\ref{lem:uniform} that
\begin{eqnarray*}
\check{\sigma}_n^2&=&L_n(\check{\theta}_n)=\frac{2\pi}{n}\sum_{j=1}^{n-1}\frac{I_Y(\lambda_j;d_0)}{H(\lambda_j;\check{\theta}_n)}+o_p(1)\\
                  &=&\frac{2\pi}{n}\sum_{j=1}^{n-1}\frac{I_X(\lambda_j)}{H(\lambda_j;\theta_0)|1-e^{i\lambda_j}|^{2m_0}}+o_p(1)=\sigma^2+o_p(1).
\end{eqnarray*}
This completes the proof.

\qed

\subsection{Lemmas}

The following lemma extends  Lemma A.2 in Velasco and Robinson
(2000) to allow conditionally heteroscedastic errors.
\begin{lemma}
\label{lem:uniform} Assume that the function
$\phi(\lambda;\theta)$ is even in $\lambda$,  periodic of period
$2\pi$ and  continuously differentiable in $\lambda$ and $\theta$
except $\lambda=0$. Further assume that there exists a $\delta\in
(0,1)$ such that for $j=1,2,\cdots,s$ and all $\theta\in\Theta$,
\begin{eqnarray}
\label{eq:phicond} &&|\phi(\lambda;\theta)|\le C
f_X^{-1}(\lambda)|\lambda|^{-\delta},~\left|\frac{\partial
\phi(\lambda;\theta)}{\partial \lambda}\right|\le
Cf_X^{-1}(\lambda)|\lambda|^{-1-\delta} \\
&&~~~~~~~~~~\mbox{and}~~\left|\frac{\partial
\phi(\lambda;\theta)}{\partial \theta_j}\right|\le
Cf_X^{-1}(\lambda)|\lambda|^{-\delta}.\nonumber
\end{eqnarray}
 Let $J_n(\theta)=2\pi
n^{-1}\sum_{j=1}^{n-1}\phi(\lambda_j;\theta)I_X(\lambda_j)$ and
$J(\theta)=\int_{0}^{2\pi}\phi(\lambda;\theta)f_X(\lambda)d\lambda$.
Suppose that
Assumptions~\ref{as:longmemo},\ref{as:utcum},\ref{as:whittle} and
\ref{as:utcov} hold. Then
\[\sup_{\theta\in\Theta}|J_n(\theta)-J(\theta)|=o_p(1)~~\mbox{as}~n\rightarrow\infty.\]
\end{lemma}
\noindent Proof of Lemma~\ref{lem:uniform}: It suffices to show
the pointwise convergence since the uniform convergence follows
from the equicontinuity argument in view of the compactness of
$\Theta$,  and differentiability of $\phi(\lambda;\theta)$ in
$\theta$.
 Let $J_n'(\theta)=2\pi
n^{-1}\sum_{j=1}^{n-1}\psi(\lambda_j;\theta) I_{u}(\lambda_j)$,
where
$\psi(\lambda;\theta)=\phi(\lambda;\theta)G(\lambda;\theta_0)$.
Hereafter in the proof, we suppress $\theta$ and write $\psi_j$
etc for $\psi(\lambda_j;\theta)$ etc. Then
\begin{eqnarray}
\label{eq:Hn} |J_n-J|\le |J_n'-\E (J_n')|+|\E
(J_n')-J|+|J_n-J_n'|.
\end{eqnarray}
 Applying Lemma~\ref{lem:covIujk},
\begin{eqnarray*}
\label{eq:Hnd} \E|J_n'-\E
(J_n')|^2&\le&Cn^{-2}\sum_{j,k=1}^{\tilde{n}}|\psi_j\psi_k||\cov(I_{uj},I_{uk})|\\
&\le&Cn^{-2}\left\{\sum_{j=1}^{\tilde{n}}\psi_j^2f_{uj}^2+n^{-1}\sum_{j\not=k=1}^{\tilde{n}}|\psi_j\psi_k|\right\}=o(1).
\end{eqnarray*}
Using the continuity of $\psi(\lambda;\theta)$ and $f_u(\lambda)$
as well as the integrability of
$\psi(\lambda;\theta)f_u(\lambda)$, we get
\begin{eqnarray*}
|\E
(J_n')-J|=\left|\frac{2\pi}{n}\sum_{j=1}^{n-1}\psi_j(f_{uj}+O(1/n))-\int_{0}^{2\pi}\psi(\lambda;\theta)f_u(\lambda)d\lambda\right|=o(1).
\end{eqnarray*}
Further, summation by parts yields
\begin{eqnarray*}
\label{eq:ixiu} &&\E|J_n-J_n'|\le
\frac{C}{n}\E\left|\sum_{j=1}^{\tilde{n}}\phi_jf_{Xj}[\tilde{I}_{Xj}-\tilde{I}_{uj}]\right|\le
\frac{C}{n}\sum_{k=1}^{\tilde{n}}\E\left|\sum_{j=1}^{k}[\tilde{I}_{Xj}-\tilde{I}_{uj}]\right|\\
&&\hspace{0.5cm}\times |\phi_k
f_{Xk}-\phi_{k+1}f_{X(k+1)}|+\frac{C}{n}\E
\left|\sum_{j=1}^{\tilde{n}}[\tilde{I}_{Xj}-\tilde{I}_{uj}]\right|
|\phi_{\tilde{n}+1}f_{X(\tilde{n}+1)}|.\nonumber
\end{eqnarray*}
By the mean value theorem, $|\phi_k
f_{Xk}-\phi_{k+1}f_{X(k+1)}|\le Cn^{-1}\lambda_k^{-1-\delta}$,
$k=1,2,\cdots,\tilde{n}$ under (\ref{eq:phicond}) and
Assumption~\ref{as:whittle}. Thus by Lemma~\ref{lem:bartlet3},
\begin{eqnarray*}
\E |J_n-J_n'|\le
\frac{C}{n^{2}}\sum_{k=1}^{\tilde{n}}\lambda_k^{-1-\delta}(k^{1/4}(1+\log
k)^{1/2}+k^{1/2}n^{-1/4})+Cn^{-3/4}\log n=o(1).
\end{eqnarray*}
Therefore the three terms on the right hand side of (\ref{eq:Hn})
are all of $o_p(1)$. This completes the proof.

\qed

\begin{lemma}
\label{lem:cltu} Under Assumptions~\ref{as:utcum},~\ref{as:utphys} and
~\ref{as:whittle},
\[T_n:=\frac{2\pi}{\sqrt{n}}\sum_{j=1}^{n-1} I_{u}(\lambda_j)\frac{\partial\log
G(\lambda_j;\theta_0)}{\partial\theta} \rightarrow_{D}
N(0,\Gamma^{(G)}(\theta_0)).\]
\end{lemma}
\noindent Proof of Lemma~\ref{lem:cltu}: Under
Assumption~\ref{as:whittle}, $\int_0^{2\pi}\partial\log
G(\lambda;\theta_0)/\partial\theta d\lambda=0$. So
\begin{eqnarray*}
\E(T_n)&=&\frac{2\pi}{\sqrt{n}}\sum_{j=1}^{n-1}(f_{uj}+O(1/n))\frac{\partial
\log G(\lambda_j;\theta_0)}{\partial\theta}\\
&=&\frac{\sigma^2\sqrt{n}}{2\pi}\left\{\int_{0}^{2\pi}\frac{\partial
\log
G(\lambda;\theta_0)}{\partial\theta}d\lambda+O(1/n)\right\}+O(n^{-1/2})=O(n^{-1/2}).
\end{eqnarray*}
Thus it suffices to show that for any $b\in \R^s$, $b'b=1$,
\begin{eqnarray}
\label{eq:Tnclt} b'\{T_n-\E (T_n)\}\rightarrow_{D} N(0, \sigma_b^2
),
\end{eqnarray}
where $\sigma_b^2=b'\Gamma^{(G)}(\theta_0)b$. A major difficulty in proving (\ref{eq:Tnclt})
 is caused by the fact that the first element of $\partial \log G(\lambda;\theta_0)/\partial\theta$ possesses a pole at zero frequency in the long memory case. We shall use a truncation argument to circumvent the problem.
 Write $b'T_n=n^{-1/2}\sum_{j=1}^{n-1}
I_{uj}\psi(\lambda_j)$, where $\psi(\lambda)=2\pi b'(\partial \log
G(\lambda;\theta_0)/\partial \theta)$. For any $c\in
 (0,1/2)$, we define two $2\pi$-periodic functions
 $\psi_1(\lambda,c)=\psi(\lambda){\bf 1}(|\lambda|\ge c)+\psi(c){\bf
 1}(|\lambda|<c)$ and
 $\psi_2(\lambda,c)=\psi(\lambda)-\psi_1(\lambda,c)$, $\lambda\in [-\pi,\pi)$. Let ${T}_{1n}(c)=n^{-1/2}\sum_{j=1}^{n-1}I_{uj}\psi_1(\lambda_j,c)$
 and
 ${T}_{2n}(c)=n^{-1/2}\sum_{j=1}^{n-1}I_{uj}\psi_2(\lambda_j,c)$.
Then (\ref{eq:Tnclt}) follows from the following two assertions:
\begin{eqnarray}
\label{eq:part1} \limsup_{c\downarrow
0}\limsup_{n\rightarrow\infty}\var({T}_{2n}(c))=0
\end{eqnarray}
and
\begin{eqnarray}
\label{eq:part2} {T}_{1n}(c)-\E \{T_{1n}(c)\} \rightarrow_{D}
N(0,\sigma_b^2(c))~\mbox{and}~\lim_{c\downarrow
0}\sigma_b^2(c)=\sigma_b^2.
\end{eqnarray}
By Lemma~\ref{lem:covIujk}, we have
\begin{eqnarray*}
&&\var(T_{2n}(c))\le\frac{C}{n}\sum_{j,k=1}^{[cn]}\cov(I_{uj},
I_{uk})(\psi(\lambda_j)-\psi(c))(\psi(\lambda_k)-\psi(c))\\
&&\le \frac{C}{n}\sum_{j=1}^{[cn]}f_{uj}^2(\psi(\lambda_j)-\psi(c))^2+\frac{C}{n^2}\sum_{j\not=k=1}^{[cn]}|f_4(\lambda_j,-\lambda_k,\lambda_k)||\psi(\lambda_j)-\psi(c)||\psi(\lambda_k)-\psi(c)|\\
&&\le
C\left\{\int_{0}^{2c\pi}f_u^2(\lambda)\psi^2(\lambda)d\lambda+\psi^2(c)c+n^{-2}\left(\sum_{j=1}^{[cn]}|\psi(\lambda_j)-\psi(c)|\right)^2\right\},
\end{eqnarray*}
which tends to zero as $c\downarrow 0$. So  (\ref{eq:part1})
holds.

We shall further approximate $T_{1n}(c)$ using techniques in Fourier analysis.
Let $d_k=(2\pi)^{-1}\int_{-\pi}^{\pi} \psi_1(\lambda,c)
e^{ik\lambda}d\lambda$ be the Fourier coefficient of
$\psi_1(\lambda,c)$. For a fixed $h\in\N$, let
$\psi_h(\lambda)=\sum_{|k|<h}(1-|k|/h)d_k e^{-ik\lambda}$ be the
Cesaro mean of the first $h$ Fourier approximations to
$\psi_1(\lambda,c)$ and
$\bar{\psi}_h(\lambda)=\psi_1(\lambda,c)-\psi_h(\lambda)$ be the
remainder. Write
$T_{1n}(c)=n^{-1/2}\sum_{j=1}^{n-1}I_{uj}\{\psi_h(\lambda_j)+\bar{\psi}_h(\lambda_j)\}$.
 It is not hard to see that
\begin{eqnarray*}
&&\limsup_{n\rightarrow\infty}
\var\left(\frac{1}{\sqrt{n}}\sum_{j=1}^{n-1}I_{uj}\bar{\psi}_h(\lambda_j)\right)=
\limsup_{n\rightarrow\infty} \frac{1}{n}\sum_{j,k=1}^{n-1}\cov(I_{uj},I_{uk})\bar{\psi}_h(\lambda_j)\bar{\psi}_h(\lambda_k)\\
&&\le
\limsup_{n\rightarrow\infty}\frac{C}{n}\left\{\sum_{j=1}^{\tilde{n}}f_{uj}^2\bar{\psi}_h^2(\lambda_j)+\frac{1}{n}\sum_{j\not=k=1}^{\tilde{n}}|f_{4}(\lambda_j,-\lambda_k,\lambda_k)\bar{\psi}_h(\lambda_j)\bar{\psi}_h(\lambda_k)|\right\}\\
&&\le C\sup_{\lambda\in
[0,2\pi]}|\bar{\psi}_h(\lambda)|^2\rightarrow
0~\mbox{as}~h\rightarrow\infty
\end{eqnarray*}
by  Fej$\acute{e}$r's theorem.

Letting $B_h=(2\pi)^{-1}(d_0,2d_1(1-1/h),\cdots,2d_{h-1}/h)'$,
$\hat{\gamma}_u(k)=n^{-1}\sum_{j=1}^{n-|k|}u_j u_{j+|k|}$ and
$\hat{\pmb{\gamma}}_u(h)=(\hat{\gamma}_u(0),\hat{\gamma}_u(1),\cdots,\hat{\gamma}_u(h-1))'$,
then
\begin{eqnarray}
\label{eq:multi} \frac{1}{\sqrt{n}}\sum_{j=1}^{n-1}[I_{uj}-\E(
I_{uj})]\psi_h(\lambda_j)&=&\frac{\sqrt{n}}{2\pi}\sum_{|k|<h}(1-|k|/h)d_k\{\hat{\gamma}_u(k)-\E
\hat{\gamma}_u(k)\}\nonumber\\
&=&\sqrt{n}B_h'[\hat{\pmb{\gamma}}_u(h)-\E
\hat{\pmb{\gamma}}_u(h)].
\end{eqnarray}

By Theorem 1 in Wu (2005), for fixed $h\in\N$,
\[\|{\cal P}_0 u_t
u_{t+h}\|\le \|u_tu_{t+h}-u_t'u_{t+h}'\|\le
C(\delta_4(t)+\delta_{4}(t+h)).\]  Then Assumption~\ref{as:utphys}
implies that  $\sum_{t=0}^{\infty}\|{\cal P}_0 u_t u_{t+h}\|\le
\infty$, which
subsequently leads to the joint asymptotic normality of
$\sqrt{n}\{\hat{\pmb{\gamma}}_u(h)-\E \hat{\pmb{\gamma}}_u(h)\}$
in view of Theorem 1(i) in Hannan (1973a) or Lemma 1 in Wu and Min (2005). Finally, it is easy to
see  that
 $\sigma^2_b(c)$ approaches  $\sigma_b^2$ as $c\downarrow 0$. Thus
 the conclusion follows.

\qed

\begin{lemma}
\label{lem:Ln}Under the assumptions in
Theorem~\ref{th:consistency}, the random variables
$L_{3n}(\theta)$ and $\check{L}_{1n}(\theta)$ defined in
(\ref{eq:L123n}) and (\ref{eq:checkln}) satisfy
\[\sup_{\theta\in\Theta_1}|L_{3n}(\theta)/\check{L}_{1n}(\theta)|=o_p(1).\]
\end{lemma}
\noindent Proof of Lemma~\ref{lem:Ln}: Note that
$\tau_j(d)=e^{i\lambda_j}\sum_{r=m_0\wedge m+1}^{m\vee
m_0}(1-e^{i\lambda_j})^{-r}Z_r$. We have
\begin{eqnarray*}
|\tau_j(d)|^2&\le&C\sum_{r=m_0\wedge m+1}^{m\vee
m_0}|1-e^{i\lambda_j}|^{-2r}Z_r^2.
\end{eqnarray*}
Since $0<C_1<K(\lambda;\theta)<C_2<\infty$
for any $\lambda$ and $\theta$, we get
\begin{eqnarray*}
\check{L}_{1n}(\theta)&\ge&\frac{C}{n}\sum_{j=1}^{n-1}|1-e^{i\lambda_j}|^{2d-2d_0}
\end{eqnarray*}
and
\begin{eqnarray*}
L_{3n}(\theta)&\le&\frac{C}{n}\sum_{r=m_0\wedge m+1}^{m\vee
m_0}\sum_{j=1}^{n-1}|1-e^{i\lambda_j}|^{2(d-r)}Z_r^2.
\end{eqnarray*}
When $m>m_0$, $Z_r^2=O_p(n^{-1})$ for
$r=m_0+1,\cdots,m$. So, uniformly in $\theta\in\Theta_1$,
\begin{eqnarray*}
\frac{L_{3n}(\theta)}{\check{L}_{1n}(\theta)}&\le&C O_p(n^{-1})\frac{\sum_{j=1}^{n-1}\sum_{r=m_0+1}^{m}|1-e^{i\lambda_j}|^{2(d-r)}}{\sum_{j=1}^{n-1}|1-e^{i\lambda_j}|^{2(d-d_0)}}=O_p(n^{-1})=o_p(1).
\end{eqnarray*}
When $m<m_0$ and $\theta\in\Theta_1$,  $m=m_0-1$, $d_{X0}<0$, $d_{X}>0$ and $Z_{m_0}^2=O_p(n^{2(d_0-m_0)})$. Therefore,
\begin{eqnarray*}
\frac{L_{3n}(\theta)}{\check{L}_{1n}(\theta)}&\le&C\frac{\sum_{j=1}^{n-1}|1-e^{i\lambda_j}|^{2(d-m_0)}O_p(n^{2(d_0-m_0)})}{\sum_{j=1}^{n-1}|1-e^{i\lambda_j}|^{2(d-d_0)}}=O_p(n^{-2\Delta})=o_p(1)
\end{eqnarray*}
uniformly in  $\theta\in\Theta_1$. The conclusion follows.
 \qed

\bigskip

Denote by $d_{\Delta}=d_0-1/2+\Delta$ for some $\Delta\in
(0,1/4)$.
\begin{lemma}
\label{lem:lower}
 Suppose that the assumptions of Theorem~\ref{th:consistency} are
satisfied. Then as $n\rightarrow\infty$, uniformly in $d\in
[a_1,d_{\Delta}]$ (if it is nonempty),
\[ L_{n}(\theta)\ge
C(1+o_p(1))J_{n}(d_{\Delta}),~~J_{n}(d_{\Delta})=\frac{1}{n}\sum_{j=1}^{\tilde{n}}(j/n)^{2d_{\Delta}}I_{Xj}|1-e^{i\lambda_j}|^{-2m_0},\]
where $C$ is a positive constant that does not depend on $\Delta$.
\end{lemma}
\noindent Proof of Lemma~\ref{lem:lower}: The proof follows the
argument in the proof of Lemma 4.2 of Abadir et al. (2007). For
the sake of completeness, we present the details here.
Note that when $d\in [a_1,d_{\Delta}]$,
\begin{eqnarray*}
L_n(\theta)&\ge&\frac{C}{n}\sum_{j=1}^{\tilde{n}}(j/n)^{2d_{\Delta}}I_Y(\lambda_j;d)\ge
C(J_{n}(d_{\Delta})-2|C_n(d)|+|B_n(d)|)
\end{eqnarray*}
for some $C>0$, where
\begin{eqnarray*}
C_n(d)&=&\frac{1}{n}\sum_{j=1}^{\tilde{n}}(j/n)^{2d_{\Delta}}w_Y(\lambda_j;d_0)\overline{\tau_j(d)}\\
B_n(d)&=&\frac{1}{n}\sum_{j=1}^{\tilde{n}}(j/n)^{2d_{\Delta}}|\tau_j(d)|^2.
\end{eqnarray*}
We shall show that $C_n(d)=o_p(1)(J_{n}(d_{\Delta})+B_n(d))$. Then
the conclusion follows since $B_n(d)\ge 0$. To this end,  let
$\delta\in (0,1/4)$ be a small fixed number. Write $C_n(d)\le
(2\pi)^{-d_0}( D_{n,1}+D_{n,2})$, where
\[D_{n,1}=n^{-1}\left|\sum_{j=1}^{\lfloor \delta n\rfloor} (j/n)^{2d_{\Delta}-d_0}v_j\overline{\tau_j(d)}\right|,~D_{n,2}=n^{-1}\left|\sum_{j=1+\lfloor \delta n\rfloor}^{\tilde{n}} (j/n)^{2d_{\Delta}-d_0}v_j\overline{\tau_j(d)}\right|\]
with
$v_j=w_X(\lambda_j)(1-e^{i\lambda_j})^{-m_0}*\lambda_j^{d_0}$.

We shall first show that $D_{n,1}\le
C_{1\delta}(1+o_p(1))(B_n(d))^{1/2}$, where the constant
$C_{1\delta}>0$ does not depend on $d$ and $n$, and
$C_{1\delta}\rightarrow 0$ as $\delta\rightarrow 0$. By the
Cauchy-Schwarz inequality,
\begin{eqnarray*}
D_{n,1}&\le&\left(n^{-1}\sum_{j=1}^{\lfloor\delta
n\rfloor}(j/n)^{2d_{\Delta}-2d_0}|v_j|^2\right)^{1/2}\left(n^{-1}\sum_{j=1}^{\tilde{n}}(j/n)^{2d_{\Delta}}|\tau_j(d)|^2\right)^{1/2}
\end{eqnarray*}
where the square of the first term is
\begin{eqnarray}
\label{eq:Dn1}
n^{-1}\sum_{j=1}^{\lfloor\delta
n\rfloor}(j/n)^{-1+2\Delta}|v_j|^2\ge\frac{C}{
n}\sum_{j=1}^{\lfloor\delta
n\rfloor}(j/n)^{-1+2\Delta}|g_j|^2\rightarrow_p
C\int_{0}^{\delta}x^{-1+2\Delta}dx=C_{1\delta}.
\end{eqnarray}
Note that the constant  $C$ in the preceding display does not
depend on $\delta$ and the convergence in (\ref{eq:Dn1}) holds
since by Lemmas~\ref{lem:bartlet1},~\ref{lem:bartlet2} and~\ref{lem:covIujk},
\begin{eqnarray*}
\frac{1}{
n}\sum_{j=1}^{\lfloor\delta
n\rfloor}(j/n)^{-1+2\Delta}(\E||g_j|^2-|h_j|^2|)&=&o(1),\\
 n^{-1}\sum_{j=1}^{\lfloor\delta
n\rfloor}(j/n)^{-1+2\Delta}\E |h_j|^2&\rightarrow& \int_0^{\delta} x^{-1+2\Delta} dx~\mbox{and}~\\
\var\left(\frac{1}{
n}\sum_{j=1}^{\lfloor\delta
n\rfloor}(j/n)^{-1+2\Delta}|h_j|^2\right)&=&o(1).
\end{eqnarray*}

Next, we show that $D_{n,2}=o_p(1)(B_n(d))^{1/2}$. Let
$B(d)=\sum_{r=m+1}^{m_0}Z_r^2$. Denote by $S_j=\sum_{l=\lfloor
\delta n\rfloor+1}^{j}v_l$ and
$t_j=(j/n)^{2d_{\Delta}-d_0}\tau_j(d)$, $j=\lfloor \delta
n\rfloor+1, \cdots,\tilde{n}$. Summation by parts implies that
\begin{eqnarray*}
D_{n,2}\le Cn^{-1}\sum_{j=\lfloor \delta
n\rfloor }^{\tilde{n}}|S_j||t_j-t_{j+1}|+|S_{\tilde{n}}t_{\tilde{n}}|.
\end{eqnarray*}
Following the same argument as in Lemma 4.2 of Abadir et al.
(2007),  the mean value theorem implies that
\[|t_j-t_{j+1}|\le C_{2\delta}j^{-1}(B(d))^{1/2},~~\mbox{and}~~|t_j|\le
C_{2\delta}(B(d))^{1/2}\]
uniformly in $d\in [a_1,d_{\Delta}]$ and
$j=\lfloor \delta n\rfloor+1,\cdots,\tilde{n}$.

So we have
\begin{eqnarray*}
D_{n,2}\le C_{2\delta}
(B(d))^{1/2}V_n,~V_n=n^{-1}\left(\sum_{j=\lfloor \delta
n\rfloor}^{n-1}|S_j|j^{-1}+|S_{\tilde{n}}|\right).
\end{eqnarray*}
 By Lemma~\ref{lem:bartlet1}, when $1\le k<j\le \tilde{n}$,
\[|\E(v_j\overline{v_k})|=\left|\frac{\sqrt{f_{Xj}}\lambda_j^{d_0}\lambda_k^{d_0}}{\sqrt{f_{Xk}}(1-e^{i\lambda_j})^{m_0}(1-e^{i\lambda_k})^{m_0}}\right||\E(g_j\overline{g_k})|\le C\log j/k\]
 and $\E(v_j\overline{v_j})\le C\E(g_j\overline{g_j})\le C(1+\log
 j/j)$. So  $\E(S_j^2)=O(j\log^2 j)$, which implies $\E(V_n)=o(1)$
 and $V_n=o_p(1)$. Finally, in view of Lemma~\ref{lem:lowerbound},
there exists an $\eta>0$, such that
\begin{eqnarray*}
B_n(d)&\ge
&\frac{C}{n}\sum_{j=1}^{\tilde{n}}(j/n)^{2m_0+1}\left|\sum_{r=m+1}^{m_0}(1-e^{i\lambda_j})^{-r}Z_r\right|^2\ge
\eta\sum_{r=m+1}^{m_0}Z_r^2=\eta B(d).
 \end{eqnarray*}
 So uniformly in $d\in [a_1,d_{\Delta}]$, $C_n(d)=o_p(1)(B_n(d))^{1/2}=o_p(1)(1+B_n(d))$. Since
 $J_n(d_{\Delta})\rightarrow_p n^{-1}\sum_{j=1}^{\tilde{n}}(j/n)^{2d_{\Delta}} f_{Xj}|1-e^{i\lambda_j}|^{-2m_0}>0$ (by Lemma~\ref{lem:uniform}),
 $C_n(d)=o_p(1)(J_n(d_{\Delta})+B_n(d))$.
 Therefore the conclusion follows. \qed

\subsection{Auxiliary Lemmas}

 We introduce the following working assumption, which holds for
 uncorrelated process $\{u_t\}$.
\begin{assumption}
\label{as:utcov} Assume that $\sum_{k\in\Z}|k\gamma_u(k)|<\infty$,
where $\gamma_u(k)={\rm cov}(u_t, u_{t+k})$.
\end{assumption}

The following three lemmas are extensions of Lemmas 6.2-6.4 in
Shao and Wu (2007b), where the results hold uniformly in
$j=1,\cdots,m$, $m=o(n)$.
\begin{lemma}
\label{lem:bartlet1}  Under Assumptions~\ref{as:longmemo} and
~\ref{as:utcov}, the following expressions hold uniformly in $1\le
k<j\le \tilde{n}$:
 \begin{eqnarray}
\label{eq:gjhj} |\E\{g_j\bar{g}_j\}-1| + |\E\{h_j\bar{h}_j\}-1|
 + |\E\{g_j\bar{h}_j\} - A_j / |A_j| |
 = O(\log j/j);
\end{eqnarray}
$\E\{g_jg_j\}=O(\log j/j)$, ~$\E\{g_j g_k\}=O(\log j/k)$,~$\E\{g_j
\bar{g}_k\}=O(\log j/k)$; \\
$\E\{h_j h_j\}=O(\log j/j)$,~$\E\{h_jh_k\}=O(\log j/k)$,
~$\E\{h_j\bar{h}_k\}=O(\log j/k)$;\\
$\E\{g_j h_j\}=O(\log j/j)$,~$\E\{g_j h_k\}=O(\log j/k)$,
~$\E\{g_j \bar{h}_k\}=O(\log j/k)$.
\end{lemma}
\noindent {Proof of Lemma~\ref{lem:bartlet1}}: The proof largely
follows the argument in Theorem 2 of Robinson (1995a), where the
uniformness is proved for $j=1,\cdots,m=o(n)$. A detailed check of
its proof shows that the argument still goes through.

\qed

\begin{lemma}
\label{lem:bartlet2}
 Suppose
Assumptions~\ref{as:longmemo} and~\ref{as:utcov} hold. Then
\begin{eqnarray}
\label{eq:ixjiuj} \E|\tilde{I}_{Xj}-\tilde{I}_{uj}|=O(j^{-1/2})
\mbox{ uniformly in } ~j=1, \cdots, \tilde{n}.
\end{eqnarray}
\end{lemma}
\noindent{Proof of Lemma~\ref{lem:bartlet2}}: It follows from the
argument of Lemma 6.3 in Shao and Wu (2007b) and the fact that
\begin{eqnarray}
\label{eq:bound1}
\int_{-\pi}^{\pi}K(\lambda-\lambda_j)\left|\frac{A(\lambda)}{A(\lambda_j)}-1\right|^2
d\lambda=O(1/j) ~\mbox{uniformly in} ~j=1,\cdots,\tilde{n}.
\end{eqnarray}
The proof of (\ref{eq:bound1}) basically repeats the argument in
Robinson's (1995b) Lemma 3 and is omitted.

\qed

\begin{lemma}
\label{lem:bartlet3}
  Suppose
Assumptions~\ref{as:longmemo},\ref{as:utcum} and~\ref{as:utcov}
hold. Then
\begin{eqnarray*}
\E\left|\sum_{j=1}^r (\tilde{I}_{Xj} - \tilde{I}_{uj})\right|
 \le C (r^{1/4}(1+\log r)^{1/2}+r^{1/2}n^{-1/4}),~r\le\tilde{n},
\end{eqnarray*}
where $C$ is a generic constant independent of $r$ and $n$.
\end{lemma}
\noindent{ Proof of Lemma~\ref{lem:bartlet3}}: The proof
repeats the argument in Lemma 6.4 of Shao and Wu (2007b), where
the two key results needed are the absolute summability of $4$-th
cumulants [i.e. Assumption~\ref{as:utcum}] and (\ref{eq:bound1}).
We omit the details. \qed

\begin{lemma}
\label{lem:covIujk} Under Assumptions~\ref{as:utcum}
and~\ref{as:utcov}, we have
\begin{eqnarray*}
{\rm cov}(I_{uj},I_{uk})={\bf 1}(j=k)[f_{uj}^2+o(1)]+{\bf
1}(j\not= k)[2\pi
n^{-1}f_4(\lambda_j,-\lambda_k,\lambda_k)+o(1/n)]
\end{eqnarray*} uniformly in $j,k=1,2,\cdots,\tilde{n}$.
\end{lemma}
\noindent { Proof of Lemma~\ref{lem:covIujk}}: Note that
\begin{eqnarray*}
\cov(I_{uj},I_{uk})&=&\E(w_{uj}w_{uk})
\E(\overline{{w}_{uj}}\overline{{w}_{uk}})+\E(w_{uj}\overline{{w}_{uk}})
\E(\overline{{w}_{uj}}w_{uk})\\
&&+\cum(w_{uj},\overline{{w}_{uj}},w_{uk},\overline{{w}_{uk}}).
\end{eqnarray*}
Under Assumption~\ref{as:utcov},  we have
\begin{eqnarray*}
\E({w}_{uj}{w}_{uk})&=&\frac{1}{2\pi
n}\sum_{t,s=1}^{n}\gamma_u(t-s)e^{it\lambda_j+is\lambda_k}=O(1/n)\\
\E(w_{uj}\overline{{w}_{uk}})&=&\frac{1}{2\pi
n}\sum_{t,s=1}^{n}\gamma_u(t-s)e^{it\lambda_j-is\lambda_k}={\bf
1}(j=k)[f(\lambda_j)+o(1)]+O(1/n).
\end{eqnarray*}
Further, Assumption~\ref{as:utcum} implies that
\begin{eqnarray*}
&&\cum(w_{uj},\overline{{w}_{uj}},w_{uk},\overline{{w}_{uk}})=\frac{1}{4\pi^2 n^2}\sum_{t_1,t_2,t_3,t_4=1}^{n}\cum(u_{t_1},u_{t_2},u_{t_3},u_{t_4})\\
&&e^{i[t_1\lambda_j-t_2\lambda_j+t_3\lambda_k-t_4\lambda_k]}\\
&&\hspace{0.5cm}=\frac{1}{4\pi^2 n^2}\sum_{h_1,h_2,h_3=1-n}^{n-1}\cum(u_0,u_{h_1},u_{h_2},u_{h_3})e^{i[-h_1\lambda_j+h_2\lambda_k-h_3\lambda_k]}\\
&&[n-1+0\wedge h_1 \wedge h_2\wedge h_3-0 \vee h_1\vee h_2 \vee
h_3]=\frac{2\pi}{n}f_4(\lambda_j,-\lambda_k,\lambda_k)+o(1/n),
\end{eqnarray*}
where we have applied the Lebesgue dominated convergence theorem
above. The conclusion follows by noting that all the results above
hold uniformly in $j,k=1,2,\cdots,\tilde{n}$.
 \qed

The following lemma is analogous to Lemma 4.3 in Abadir et al.
(2007) and our argument seems simpler.

\begin{lemma}
\label{lem:lowerbound} Let $q\ge 0$ be a fixed integer. Then there
exists $\eta>0$ such that, as $n\rightarrow\infty$, uniformly in
$a_0,\cdots,a_q\in\R$,
\begin{eqnarray}
\label{eq:lowerbound}
n^{-1}\sum_{j=1}^{\tilde{n}}(j/n)^{2q+1}\left|\sum_{r=0}^{q}(1-e^{i\lambda_j})^{-r}a_r\right|^2\ge
\eta\sum_{r=0}^{q}a_r^2.
\end{eqnarray}
\end{lemma}
\noindent Proof of Lemma~\ref{lem:lowerbound}: Write  the
left-hand side of (\ref{eq:lowerbound}) as
$\sum_{r,s,=0}^{q}a_ra_sD_n(r,s)$, where
\[D_n(r,s)=n^{-1}\sum_{j=1}^{\tilde{n}}(j/n)^{2q+1}\{(1-e^{i\lambda_j})^{-r}(1-e^{-i\lambda_j})^{-s}\}.\]
Note that $D_n(r,s)\rightarrow D(r,s)$ for $r,s=0,\cdots,q$, where
\[D(r,s)=\frac{1}{2}\int_0^{1/2} \frac{x^{2q+1}}{|1-e^{i2\pi x}|^{2q}}(1-e^{i 2\pi x})^{q-r} (1-e^{-i 2\pi x})^{q-s}dx.\]
It is easy to see that $D=(D(r,s))_{r,s,=0,\cdots,q}$ is a
positive semidefinite Hermitian matrix, so all the eigenvalues are
real and nonnegative. We proceed to show that no eigenvalues are
zero. Suppose that there exists a vector $(G_0,G_1,\cdots,G_q)'$
such that $\sum_{s=0}^{q}D(r,s)G_s=0$ for $r=0,\cdots,q$. Let
$G(x)=\sum_{s=0}^{q}(1-e^{-i 2\pi x})^{q-s}G_s$. Then
\[\int_{0}^{1/2}\frac{x^{2q+1}}{|1-e^{i2\pi x}|^{2q}}(1-e^{i 2\pi x})^{q-r} G(x)dx=0,~r=0,\cdots,q. \]
Thus we get
\begin{eqnarray*}
\int_{0}^{1/2}\frac{x^{2q+1}}{|1-e^{i2\pi x}|^{2q}}|G(x)|^2dx&=&\int_0^{1/2}\frac{x^{2q+1}}{|1-e^{i2\pi x}|^{2q}}G(x)\overline{G(x)}dx\\
&=&\sum_{r=0}^{q}G_r \int_0^{1/2}\frac{x^{2q+1}}{|1-e^{i2\pi x}|^{2q}}G(x)(1-e^{i2\pi x})^{q-r}dx=0,
\end{eqnarray*}
which implies that $G(x)=0$ for almost all $x\in [0,1/2)$.
Consequently,  $G_s=0$, $s=0,1,\cdots,q$. Let $2\eta>0$ be the
smallest eigenvalue of $D$. Then for large enough $n$, the
eigenvalues of $D_n=(D_n(r,s))_{r,s=0,\cdots,q}$ are no smaller than $\eta$. This
completes the proof.

\qed

\bigskip

 \centerline{REFERENCES}

\bigskip
\par\noindent\hangindent2.3em\hangafter 1
{Abadir, K. M.}, {W. Distaso} \& {L. Giraitis} (2007)
Nonstationarity-extended local Whittle estimation. {\it Journal of
Econometrics} 141, 1353-1384.

\par\noindent\hangindent2.3em\hangafter 1
{Baillie, R. T.}, {C. F. Chung} \& {M. A. Tieslau} (1996)
Analyzing inflation by the fractionally integrated ARFIMA-GARCH
model.  {\it Journal of Applied Econometrics} 11, 23-40.

\par\noindent\hangindent2.3em\hangafter 1
{Beran, J.} (1995)  Maximum likelihood estimation of the
differencing parameter for invertible short and long memory
autoregressive integrated moving average models {\it Journal of
the Royal Statistical Society, B.} 57, 659-672.

\par\noindent\hangindent2.3em\hangafter 1
{Bollerslev, T.} (1986) Generalized autoregressive conditional
heteroscedasticity. {\it Journal of Econometrics} 31, 307-327.

\par\noindent\hangindent2.3em\hangafter 1
{Chiu, S. T.} (1988) Weighted least squares estimators on the
frequency domain for the parameters of a time series. {\it Annals
of Statistics}, 16, 1315-1326.

\par\noindent\hangindent2.3em\hangafter 1
{Dahlhaus, R.} (1989) Efficient parameter estimation for
self-similar processes. {\it Annals of Statistics} 17, 1749-1766.

\par\noindent\hangindent2.3em\hangafter 1
{Davidson, J.} (2004) Moment and memory properties of linear conditional heteroscedasticity models, and a new model. {\it
 Journal of Business and Economics Statistics}  22,  16-29.

\par\noindent\hangindent2.3em\hangafter 1
{Ding, Z.}, {C. Granger} \& {R. Engle} (1993) A long memory
property of stock market returns and a new model. {\it Journal of
Empirical Finance} 1, 83-106.

\par\noindent\hangindent2.3em\hangafter 1
{Doukhan, P.}, { G. Oppenheim} \& { M. S. Taqqu} (2003) {\it
Theory and Applications of Long-range Dependence}, Birkha\"user,
Boston.

\par\noindent\hangindent2.3em\hangafter 1
{Elek, P.} \& {L. M\'{a}rkus} (2004) A long range dependent
model with nonlinear innovations for simulating daily river flows.
{\it Natural Hazards and Earth System Sciences}, 4, 277-283.

\par\noindent\hangindent2.3em\hangafter 1
{Fan, J.} \& {Q. Yao} (2003). {\it Nonlinear Time Series:
Nonparametric and Parametric Methods}. Springer-Verlag, New York.

\par\noindent\hangindent2.3em\hangafter 1
{Fox, R.} \& {M. S. Taqqu} (1986) Large-sample properties of
parameter estimates for strongly dependent stationary Gaussian
time series. {\it Annals of Statistics} 14, 517-532.

\par\noindent\hangindent2.3em\hangafter 1
{Giraitis, L.} \& {P. M. Robinson} (2001) Whittle estimation of
ARCH models. {\it Econometric Theory} 17, 608-631.

\par\noindent\hangindent2.3em\hangafter 1
{Giraitis, L.} \& {D. Surgailis} (1990) A central limit theorem
for quadratic forms in strongly dependent random variables and its
application to asymptotic normality of Whittle's estimate. {\it
Probability Theory and Related Fields} 86, 87-104.

\par\noindent\hangindent2.3em\hangafter 1
{Giraitis, L.} \& {D. Surgailis} (2002) ARCH-type bilinear models with
double long memory. {\it Stochastic Processes and Their Applications}
 100, 275-300.

\par\noindent\hangindent2.3em\hangafter 1
{Giraitis, L.} \& {M. S. Taqqu} (1999)  Whittle estimator for
non-Gaussian long-memory time series. {\it Annals of Statistics}
27, 178-203.

\par\noindent\hangindent2.3em\hangafter 1
{Gray, H. L.}, {N.-F. Zhang} \& {W. A. Woodward}  (1989) On
generalized fractional processes. {\it Journal of Time Series
Analysis} 10, 233-257.

\par\noindent\hangindent2.3em\hangafter 1
{Hannan, E. J.} (1973a) Central limit theorems for time series
regression. {\it Z. Wahrsch. Verw. Geb} 26, 157-170.

\par\noindent\hangindent2.3em\hangafter 1
{Hannan, E. J.} (1973b) The asymptotic theory of linear time
series models. {\it Journal of Applied Probability} 10, 130-145.

\par\noindent\hangindent2.3em\hangafter 1
{Hauser, M. A.} \& {R. M. Kunst} (1998a) Fractionally integrated
models with ARCH errors: with an application to the Swiss 1-month
euromarket interest rate. {\it Review of Quantitative Finance and
Accounting} 10, 95-113.

\par\noindent\hangindent2.3em\hangafter 1
{Hauser, M. A.} \&  {R. M. Kunst} (1998b)  Forecasting
high-frequency financial data with the ARFIMA-ARCH model. {\it
Journal of Forecasting} 20, 501-518.

\par\noindent\hangindent2.3em\hangafter 1
{Hosoya, Y.} (1997) A limit theory for long-range dependence and
statistical inference on related models.  {\it Annals of
Statistics} 25, 105-137.

\par\noindent\hangindent2.3em\hangafter 1
{Keenan, D. M.} (1987) Limiting behavior of functionals of
higher-order sample cumulant spectra. {\it Annals of Statistics},
15, 134-151.

\par\noindent\hangindent2.3em\hangafter 1
{Koopman, S. J.}, {M. Oohs} \& {M. A. Carnero} (2007) Periodic
seasonal Reg-ARFIMA-GARCH models for daily electricity spot
prices. {\it Journal of American Statistical Association} 102,
16-27.

\par\noindent\hangindent2.3em\hangafter 1
{Lien, D.} \& {Y. K. Tse}  (1999) Forecasting the Nikkei spot
index with fractional cointegration. {\it Journal of Forecasting}
18, 259-273.

\par\noindent\hangindent2.3em\hangafter 1
{Ling, S.} \& {W. K. Li} (1997) On fractionally integrated
autoregressive moving-average time series models with conditional
heteroscedasticity. {\it Journal of the American Statistical
Association} 92, 1184-1194.

\par\noindent\hangindent2.3em\hangafter 1
{Ling, S.} \& {M. McAleer} (2002) Necessary and sufficient moment conditions for the
GARCH(r, s) and asymmetric power GARCH(r, s) models. {\it Econometric Theory} 18, 722-729.

\par\noindent\hangindent2.3em\hangafter 1
{Marinucci, D.} \& {P. M. Robinson} (1999)
Alternative forms of fractional Brownian motion.   {\it Journal of
Statistical Planning and Inference}  80, 111-122.

\par\noindent\hangindent2.3em\hangafter 1
{Mayoral, L.} (2007)  Minimum distance estimation of stationary
and non-stationary ARFIMA processes.  {\it Econometrics Journal}
10, 124-148.

\par\noindent\hangindent2.3em\hangafter 1
{Nelson, D. B.} (1991) Conditional heteroskedasticity in asset
returns: a new approach. {\it Econometrica} 59, 347-370.

\par\noindent\hangindent2.3em\hangafter 1
{Phillips, P. C. B.} (1999) Discrete Fourier transforms of
fractional processes. Technical Report. Yale University.

\par\noindent\hangindent2.3em\hangafter 1
{Phillips, P. C. B.} \& {K. Shimotsu} (2004)  Local
Whittle estimation in nonstationary and unit root cases. {\it
Annals of Statistics} 32, 656-692.

\par\noindent\hangindent2.3em\hangafter 1
{Robinson, P. M.} (1995a) Log-periodogram regression of time
series with long range dependence. {\it Annals of Statistics} 23,
1048-1072.

\par\noindent\hangindent2.3em\hangafter 1
{Robinson, P. M.} (1995b) Gaussian semiparametric estimation of
long range dependence.  {\it Annals of Statistics} 23, 1630-1661.

\par\noindent\hangindent2.3em\hangafter 1
{Robinson, P. M.} (2003) {\it Time Series with Long Memory},
Oxford University Press.

\par\noindent\hangindent2.3em\hangafter 1
{Robinson, P. M.} (2005) The distance between rival
nonstationary fractional processes.  {\it Journal of Econometrics}
128, 283-300.

\par\noindent\hangindent2.3em\hangafter 1
{Shao, X.} \& {W. B. Wu} (2007a) Asymptotic spectral theory for
nonlinear time series. {\it Annals of Statistics} 35, 1773-1801.

\par\noindent\hangindent2.3em\hangafter 1
{Shao, X.} \& {W. B. Wu} (2007b) Local Whittle estimation of
fractional integration for nonlinear processes. {\it Econometric
Theory} 23, 899-929.

\par\noindent\hangindent2.3em\hangafter 1
{Shimotsu, K.} \& {P. C. B. Phillips} (2006) Local
Whittle estimation of fractional integration and some of its
variants. {\it Journal of Econometrics} 103, 209-233.

\par\noindent\hangindent2.3em\hangafter 1
{Subba Rao, T.} \& {M. M. Gabr} (1984). {\it An Introduction to
Bispectral Analysis and Bilinear Time Series Models}. Lecture
Notes in Statistics, 24. New York: Springer-Verlag.

\par\noindent\hangindent2.3em\hangafter 1
{Taniguchi, M.} (1982) On estimation of the integrals of the 4th
order cumulant spectral density. {\it Biometrika}, 69, 117-122.

\par\noindent\hangindent2.3em\hangafter 1
{Tong, H.} (1990). {\it Non-linear Time Series: A Dynamical System
Approach.} Oxford University Press.

\par\noindent\hangindent2.3em\hangafter 1
{Velasco, C.} \& { P. M. Robinson} (2000) Whittle pseudo-maximum
likelihood estimation for nonstationary time series. {\it Journal
of the American Statistical Association} 95, 1229-1243.

\par\noindent\hangindent2.3em\hangafter 1
{Walker, A. M.} (1964) Asymptotic properties of least-squares
estimates of parameters of the spectrum of a stationary
non-deterministic time-series. {\it Journal of  Australian Mathematical Society} 4,
363-384.

\par\noindent\hangindent2.3em\hangafter 1
{Wu, W. B.}, 2005. Nonlinear system theory: another look at
dependence. {\it Proceedings of National Academy of Science, USA.} 102, 14150-14154.

\par\noindent\hangindent2.3em\hangafter 1
{Wu, W. B.} \& {M. Min} (2005).  On linear processes with
dependent innovations.  {\it Stochastic Processes and Their Applications}  115, 939-958.

\par\noindent\hangindent2.3em\hangafter 1
{Wu, W. B.} \& {X. Shao} (2004) Limit theorems for iterated random
functions. {\it Journal of Applied Probability} 41, 425-436.

\par\noindent\hangindent2.3em\hangafter 1
{Zaffaroni, P.} \& {B. d'Italia} (2003) Gaussian inference on
certain long-range dependent volatility models. {\it Journal of
Econometrics}, 115, 199-258.

\newpage

\begin{footnotesize}
\begin{center}
\begin{tabular}{c|cccccc}
\hline  \hline
(a) $n=200$&d&\vline \vline \vline&W-1&W-3&W-4&EW\\
\hline
 &-0.4&\vline \vline \vline&0.0061&-0.0920&-0.2100&-0.0280\\
&0.4&\vline \vline \vline&-0.0115&-0.0829&-0.2054&-0.0453\\
&0.6&\vline \vline \vline&-0.0429&-0.0759&-0.1995&-0.0208\\
&0.9&\vline \vline \vline&-0.0099&-0.0595&-0.1914&-0.0461\\
$\mbox{Bias}$&1.1&\vline \vline \vline&-0.1839&-0.0427&-0.1841&-0.0489\\
&1.4&\vline \vline \vline&-0.4393&0.0010&-0.1707&-0.0453\\
&1.5&\vline \vline \vline&-0.5469&0.0724&-0.1658&-0.0042\\
&2.0&\vline \vline \vline&-0.9970&0.1549&-0.1369&-0.0482\\
&2.4&\vline \vline \vline&-1.4237&0.3539&-0.1055&-0.0453\\
\hline\hline
&-0.4&\vline \vline \vline&0.9910&3.2282&12.0725& 1.0767\\
&0.4&\vline \vline \vline&0.9853&3.0389&11.854&1.2049\\
&0.6&\vline \vline \vline&1.0211&2.8958&11.359&0.8644\\
&0.9&\vline \vline \vline&1.4808&2.5940&11.002&1.1459\\
$\mbox{MSE}\times 100$&1.1&\vline \vline \vline&5.7039&2.3579&10.659&1.1765\\
&1.4&\vline \vline \vline&24.397&2.1186&10.160&1.2049\\
&1.5&\vline \vline \vline&32.610&2.9531&10.040&0.9486\\
&2.0&\vline \vline \vline&103.19&5.1102&9.7201&1.1650\\
&2.4&\vline \vline \vline&204.77&16.6550&9.4931&1.2049\\
\hline\hline
\end{tabular}
\end{center}

\begin{center}
\begin{tabular}{c|cccccc}
\hline  \hline
(b) $n=512$&d&\vline \vline \vline&W-1&W-3&W-4&EW\\
\hline
 &-0.4&\vline \vline \vline&0.0094&-0.0346&-0.0689&-0.0041\\
&0.4&\vline \vline \vline&-0.0007&-0.0316&-0.0673&-0.0139\\
&0.6&\vline \vline \vline&0.0026&-0.0288&-0.0657&-0.0034\\
&0.9&\vline \vline \vline&-0.0385&-0.0220&-0.0622&-0.0158\\
$\mbox{Bias}$&1.1&\vline \vline \vline&-0.1424&-0.0145&-0.0591&-0.0174\\
&1.4&\vline \vline \vline&-0.3695&0.0076&-0.0532&-0.0139\\
&1.5&\vline \vline \vline&-0.4894&0.0491&-0.0509&0.0104\\
&2.0&\vline \vline \vline&-0.9775&0.0948&-0.0361&-0.0170\\
&2.4&\vline \vline \vline&-1.3934&0.2474&-0.0199&-0.0139\\
\hline\hline
&-0.4&\vline \vline \vline&0.3659&0.8356&2.1561& 0.3599\\
&0.4&\vline \vline \vline&0.3208&0.8156&2.1719 &0.3517\\
&0.6&\vline \vline \vline&0.3226&0.7961&2.1558&0.3430\\
&0.9&\vline \vline \vline&0.4886&0.7547&2.1161&0.3545\\
$\mbox{MSE}\times 100$&1.1&\vline \vline \vline&2.5569&0.7203&2.0792&0.3552\\
&1.4&\vline \vline \vline&14.456&0.6973&2.0101&0.3517\\
&1.5&\vline \vline \vline&24.656&1.1706&1.9841&0.3669\\
&2.0&\vline \vline \vline&96.696&1.9106&1.8475&0.3560\\
&2.4&\vline \vline \vline&195.12&8.4293&1.7640&0.3517\\
\hline\hline
\end{tabular}
\end{center}

\begin{table}
 \caption{The bias and mean squared error (MSE) of $\hat{d}$ for the FARIMA(2, d, 0) model with GARCH(1,1) innovation~(\ref{eq:garch4}) when (a). $n=200$ and (b) $n=512$.} \label{tb:table1}
\end{table}

\newpage
\begin{center}
\begin{tabular}{c|cccccc}
\hline  \hline
(a) $n=200$&d&\vline \vline \vline&W-1&W-3&W-4&EW\\
\hline
 &-0.4&\vline \vline \vline&-0.0072&0.0400&0.0925&0.0106\\
&0.4&\vline \vline \vline&-0.0066&0.0357&0.0903&0.0118\\
&0.6&\vline \vline \vline&-0.0142&0.0326&0.0873&0.0051\\
&0.9&\vline \vline \vline&-0.0745&0.0254&0.0845&0.0162\\
$\mbox{Bias}$&1.1&\vline \vline \vline&-0.2176&0.0181&0.0816&0.0158\\
&1.4&\vline \vline \vline&-0.4684&-0.0006&0.0764&0.0117\\
&1.5&\vline \vline \vline&-0.5458&-0.0305&0.0746&0.0062\\
&2.0&\vline \vline \vline&-0.6195&-0.0620&0.0643&0.0163\\
&2.4&\vline \vline \vline&-0.6155&-0.1172&0.0526&0.0118\\
\hline\hline
&-0.4&\vline \vline \vline&0.7545&1.7357&4.6639&0.7599\\
&0.4&\vline \vline \vline&0.7668&1.7158&4.6111&0.7673\\
&0.6&\vline \vline \vline&0.8337&1.6925&4.4636&0.6759\\
&0.9&\vline \vline \vline&2.1480&1.6399&4.4214&0.7584\\
$\mbox{MSE}\times 100$&1.1&\vline \vline \vline&8.8843&1.5993&4.3566&0.7604\\
&1.4&\vline \vline \vline&27.606&1.5721&4.2666&0.7670\\
&1.5&\vline \vline \vline&32.433&1.7772&4.2542&0.5652\\
&2.0&\vline \vline \vline&40.138&2.0674&4.2813&0.7592\\
&2.4&\vline \vline \vline&38.784&2.9361&4.2722&0.7672\\
\hline\hline
\end{tabular}
\end{center}
\begin{center}
\begin{tabular}{c|cccccc}
\hline  \hline
(b) $n=512$ &d&\vline \vline \vline&W-1&W-3&W-4&EW\\
\hline
 &-0.4&\vline \vline \vline&-0.0067&0.0140&0.0327&0.0003\\
&0.4&\vline \vline \vline&-0.0057&0.0126&0.0320&0.0014\\
&0.6&\vline \vline \vline&-0.0115&0.0113&0.0313&-0.0002\\
&0.9&\vline \vline \vline&-0.0779&0.0082&0.0298&0.0044\\
$\mbox{Bias}$&1.1&\vline \vline \vline&-0.2859&0.0047&0.0285&0.0043\\
&1.4&\vline \vline \vline&-0.5778&-0.0056&0.0259&0.0015\\
&1.5&\vline \vline \vline&-0.6147&-0.0245&0.0249&-0.0034\\
&2.0&\vline \vline \vline&-0.6599&-0.0443&0.0185&0.0046\\
&2.4&\vline \vline \vline&-0.6501&-0.0951&0.0115&0.0015\\
\hline\hline
&-0.4&\vline \vline \vline&0.2856&0.5727&1.1368&0.2810\\
&0.4&\vline \vline \vline&0.2702&0.5707&1.1508&0.2672\\
&0.6&\vline \vline \vline&0.2901&0.5680&1.1502&0.2733\\
&0.9&\vline \vline \vline&1.5366&0.5618&1.1458&0.2726\\
$\mbox{MSE}\times 100$&1.1&\vline \vline \vline&11.753&0.5577&1.1406&0.2691\\
&1.4&\vline \vline \vline&35.090&0.5700&1.1291&0.2672\\
&1.5&\vline \vline \vline&38.582&0.7093&1.1245&0.2275\\
&2.0&\vline \vline \vline&43.940&0.8519&1.0990&0.2709\\
&2.4&\vline \vline \vline&42.661&1.6801&1.0830&0.2672\\
\hline\hline
\end{tabular}
\end{center}
\begin{table}
 \caption{The bias and mean squared error (MSE) of $\hat{\phi_1}$ for the FARIMA(2, d, 0) model with GARCH(1,1) innovation~(\ref{eq:garch4})  when (a). $n=200$ and (b) $n=512$.}
 \label{tb:table2}
\end{table}

\newpage

\begin{center}
\begin{tabular}{c|cccccc}
\hline  \hline
(a) $n=200$&d&\vline \vline \vline&W-1&W-3&W-4&EW\\
\hline
&-0.4&\vline \vline \vline&0.0078&-0.0013&-0.0112&0.0077\\
&0.4&\vline \vline \vline&0.0112&-0.0007&-0.0109&0.0105\\
&0.6&\vline \vline \vline&0.0217&-0.0001&-0.0100&0.0077\\
&0.9&\vline \vline \vline&0.1256&0.0013&-0.0091&0.0067\\
$\mbox{Bias}$&1.1&\vline \vline \vline&0.3226&0.0027&-0.0082&0.0073\\
&1.4&\vline \vline \vline&0.5511&0.0067&-0.0067&0.0105\\
&1.5&\vline \vline \vline&0.6015&0.0133&-0.0061&0.0063\\
&2.0&\vline \vline \vline&0.6081&0.0236&-0.0031&0.0068\\
&2.4&\vline \vline \vline&0.6158&0.0629&0.0005&0.0105\\
\hline\hline
&-0.4&\vline \vline \vline&0.4422&0.8304&1.1368&0.4439\\
&0.4&\vline \vline \vline&0.4715&0.8302&1.2365&0.4750\\
&0.6&\vline \vline \vline&0.5650&0.8330&1.2421&0.4392\\
&0.9&\vline \vline \vline&3.4191&0.8408&1.2340&0.4445\\
$\mbox{MSE}\times 100$&1.1&\vline \vline \vline&14.553&0.8495&1.2369&0.4497\\
&1.4&\vline \vline \vline&32.547&0.8735&1.2481&0.4744\\
&1.5&\vline \vline \vline&36.947&0.9145&1.2523&0.3709\\
&2.0&\vline \vline \vline&37.185&1.0062&1.2853&0.4467\\
&2.4&\vline \vline \vline&38.044&1.4756&1.3220&0.4749\\
\hline\hline
\end{tabular}
\end{center}

\begin{center}
\begin{tabular}{c|cccccc}
\hline  \hline
(b) $n=512$&d&\vline \vline \vline&W-1&W-3&W-4&EW\\
\hline
 &-0.4&\vline \vline \vline&0.0040&0.0029&0.0019&0.0041\\
&0.4&\vline \vline \vline&0.0061&0.0030&0.0020&0.0061\\
&0.6&\vline \vline \vline&0.0128&0.0032&0.0021&0.0042\\
&0.9&\vline \vline \vline&0.1161&0.0035&0.0024&0.0037\\
$\mbox{Bias}$&1.1&\vline \vline \vline&0.3559&0.0039&0.0026&0.0040\\
&1.4&\vline \vline \vline&0.5636&0.0051&0.0031&0.0061\\
&1.5&\vline \vline \vline&0.5905&0.0076&0.0033&0.0044\\
&2.0&\vline \vline \vline&0.5969&0.0113&0.0045&0.0038\\
&2.4&\vline \vline \vline&0.6009&0.0354&0.0059&0.0061\\
\hline\hline
&-0.4&\vline \vline \vline&0.1729&0.3392&0.5846&0.1734\\
&0.4&\vline \vline \vline&0.1793&0.3393&0.5847&0.1800\\
&0.6&\vline \vline \vline&0.2090&0.3394&0.5852&0.1730\\
&0.9&\vline \vline \vline&2.6947&0.3399&0.5863&0.1730\\
$\mbox{MSE}\times 100$&1.1&\vline \vline \vline&16.354&0.3406&0.5872&0.1742\\
&1.4&\vline \vline \vline&32.879&0.3438&0.5891&0.1801\\
&1.5&\vline \vline \vline&35.414&0.3519&0.5899&0.1455\\
&2.0&\vline \vline \vline&35.700&0.3631&0.5949&0.1735\\
&2.4&\vline \vline \vline&36.174&0.5482&0.6009&0.1801\\
\hline\hline
\end{tabular}
\end{center}
\begin{table}
 \caption{The bias and mean squared error (MSE) of $\hat{\phi_2}$ for the FARIMA(2, d, 0) model with GARCH(1,1) innovation~(\ref{eq:garch4})  when (a) $n=200$ and (b) $n=512$.} \label{tb:table3}
\end{table}

\end{footnotesize}

\end{document}